\documentclass[aps,nofootinbib,prd]{revtex4}
\usepackage{graphicx}
\usepackage[dvipsnames]{xcolor}
\usepackage{amsmath}
\usepackage{amssymb}
\usepackage{amsfonts}
\usepackage{ulem}
\usepackage{hyperref}

\newcommand{\mpl}{M_{\rm{P}}}

\newcommand{\ESFM}{Departamento de F\'{\i}sica, Escuela Superior de
F\'{\i}sica y Matem\'aticas del Instituto Polit\'ecnico Nacional \\
Unidad Adolfo L\'opez Mateos, Edificio 9, 07738 Ciudad de M\'exico, Mexico}

\newcommand{\Catedra}{Consejo
  Nacional de Ciencia y Tecnolog\'{\i}a.  Av.  Insurgentes Sur 1582,
  03940, Ciudad de M\'exico, M\'exico}

\newcommand{\Cinvestav}{Departamento de F\'{\i}sica, Centro de
  Investigaci{\'o}n y de Estudios Avanzados del IPN\\ Apdo. Postal
  14-740 07000 Ciudad de M\'exico, Mexico}

\begin{document}

\title{Perturbations and stability conditions of k-essence and kinetic gravity
  braiding models in two-field measure theory}

\author{R. Cordero~$^1$}\email{rcorderoe@ipn.mx}
\author{J. De-Santiago~$^{2, 3}$}\email{Josue.desantiago@cinvestav.mx}
\author{O. G. Miranda~$^3$}\email{omr@fis.cinvestav.mx}
\author{M. Serrano-Crivelli~$^1$}\email{mserranoc0800@alumno.ipn.mx}
\affiliation{$^1$~\ESFM}
\affiliation{$^2$~\Catedra}
\affiliation{$^3$~\Cinvestav}

\begin{abstract}
We study cosmological perturbations for k-essence and kinetic gravity braiding models in the context of
the two-field measure theory (TMT). Considering scalar perturbations and the uniform field gauge, we obtain the sound speed of the fields and present a stability analysis by means of the kinetic matrix and the mass eigenvalues. For k-essence models, in the two-field measure theory, the speed of propagation of the field is modified completely due to the new measure field and it gives rise to crucial differences with respect to the case without new measure. The stability analysis gives a physical viable model for the Universe. For the kinetic gravity braiding models in the two-field measure theory we get that, in general, the speed of perturbations is equal to the speed of light which is a consequence of the properties of the new measure field. In the later case, there is always a ghost field. Furthermore, we calculate general expressions for the mass eigenvalues and find, for an explicit example, the existence of tachyonic instabilities.

\end{abstract}


\maketitle
\section{Introduction}
The discovery of the Universe's accelerating expansion~\cite{Riess:1998cb,Perlmutter:1998np} became one of the most
interesting current
problems in cosmology~\cite{Sahni:2004ai,Alam:2004jy,Sahni:2006pa,Feng:2004ad,Durrer:2007re,Bamba:2012cp}. To explain this behaviour, scientists have proposed many different dark
energy models that can be classified into two kinds: the matter and gravity-modified models (see, for example,
\cite{Bianchi:2010uw,Ratra:1987rm,Copeland:2006wr,Linder:2007wa,Deffayet:2001pu,Shtanov:2002mb,Amendola:2015ksp,Aviles:2018saf}).

A recent alternative explanation to the
observations of an accelerating expanding Universe is the Two Field
Measure Theory (TMT)~\cite{Guendelman:2012gg,Ansoldi:2012pi}. This
model includes a new measure, $\Phi$, in its action, besides the usual general relativity measure ($\sqrt{-g}$).
Different actions have been analyzed in
this context~\cite{Guendelman:2013ke,Guendelman:2015rea,Guendelman:2015jii,Cordero:2019mze,Bensity:2020sfu,Benisty:2018fja}.
Recently, it has been argued that this kind of
formalism is useful in string and brane theories to generate a string tension that
appears dynamically~\cite{Guendelman:2021nve,Guendelman:2021bbr,Vulfs:2019xtb}. One of the more interesting properties is that for general k-essence models, this proposal unifies dark matter and dark energy~\cite{Guendelman:2012gg,Ansoldi:2012pi}. Moreover, the role of fermionic matter in TMT has also been studied and it was found that its incorporation may lead to a dynamically generated dark energy, for instance, through the neutrino sector~\cite{Guendelman:2006ji,Guendelman:2012vc}. However, to test its viability as an acceptable cosmological model, it is necessary to study the perturbations of these models to see if they evolve according to the observations without developing ghosts or exponential growth. For example, there exist ghost modes in the self-accelerating branch of the Dvali-Gabadadze-Porrati cosmological model ~\cite{Luty:2003vm}, in a nonlinear massive gravity \cite{Boulware:1972zf} and in the linear massive gravity with a de Sitter background spacetime \cite{Higuchi:1986py}. Several aspects of perturbations in TMT models have been studied in the k-essence context in~\cite{Guendelman:2008sv, Guendelman:2012gg} by a different approach to the presented here.

Some of the most studied models with the possibility to drive the late-time accelerated expansion of the Universe lie in the category of Horndeski's models~\cite{horndeski1974second},
which are scalar-tensor theories with second-order equations of motion.
From the classification found in \cite{Deffayet:2011gz}, we will focus on the first two types of Lagrangians, the k-essence~\cite{Armendariz-Picon:1999hyi,Armendariz-Picon:2000ulo,Armendariz-Picon:2000nqq,Melchiorri:2002ux,Chiba:1999ka,Chiba:2002mw,Chimento:2003zf,Chimento:2003ta,De-Santiago:2011aka,De-Santiago:2012ibi,Cordero:2016bxt} and the kinetic braiding models~\cite{Deffayet:2010qz}, which are not excluded by recent observations of gravitational waves~\cite{LIGOScientific:2017vwq,LIGOScientific:2017ync,LIGOScientific:2017zic,Kase:2018aps,Kobayashi:2019hrl}. 
Other physical preferences could be extracted from the study of the Hubble constant tension in Hordenski's  models which can be found in \cite{Banerjee:2020xcn,Lee:2022cyh}.
The k-essence models are characterized by the action\footnote{This kind of models were proposed initially for inflation \cite{Armendariz-Picon:1999hyi, Garriga:1999vw}.}
\begin{equation}
    S = \int K(\varphi,X) \sqrt{-g} d^4x \,,
\end{equation}
where the Lagrangian $K$  is
a general function of the k-essence field $\varphi$ and its kinetic term $X=-1/2 g^{\mu\nu}\partial _\mu \varphi \partial _\nu \varphi$, where $g_{\mu \nu}$ is
the metric of the spacetime with signature $(-+++)$. Later on, it was realized that this model belongs to a wider class of models called Galileons \cite{Nicolis:2008in,deRham:2010eu,Goon:2010xh,Goon:2011qf}.
In particular, some k-essence models can provide a mechanism to produce a unified description of dark energy and dark matter ~\cite{Scherrer:2004au,De-Santiago:2011aka,De-Santiago:2012ibi}.
The second very interesting subclass of Horndeski's models is the so-called kinetic gravity braiding models. These models have an action of the form
\begin{equation}
    S = -\int {G_3}(\varphi,X) \square \varphi \sqrt{-g} d^4x \,,
\end{equation}
that includes a D'Alambertian operator and a general function of the field and its
kinetic term~\cite{Deffayet:2010qz}. \footnote{The notation $G_3$ for the Lagrangian comes from the
classification of the generalized Galileon models~\cite{Deffayet:2011gz}.} In this model, the pressure depends on second-order derivatives of the scalar fields, and the expression for the energy density
contains the Hubble parameter. These characteristics produce exciting features analyzed in Refs.~\cite{Pujolas:2011he,Kimura:2010di,Kimura:2011td,Maity:2012dx}.

In this paper, we study the cosmological scalar perturbations of k-essence and kinetic gravity braiding models in the context of two field measure theory. To do this we perturb
the action following the procedure in~\cite{DeFelice:2011bh}. In particular, we obtain propagation speed of the fields, the kinetic matrix and the mass eigenvalues for the kinetic gravity braiding model in TMT. We illustrate our calculations for a particular braiding field.

The paper is organized as follows. In Section~\ref{section:TMT}, we develop the main ingredients of TMT.
In Section~\ref{section:kessence}, we compute the scalar perturbations and the speed of sound of the field for k-essence  models and compare them with the case when the TMT is included.
In Section~\ref{section:braiding}, we calculate the perturbations of the kinetic gravity braiding model in TMT and the kinetic matrix. In particular, we present the procedure to obtain the mass eigenvalues.
Finally, we give our conclusions.
Additionally, as a guide to the reader, we include table~\ref{tab1} with the symbols used in this work.

\begin{table}
    \begin{tabular}{ |c|c|c| }
 \hline
 $\varphi$   & Horndeski's field.  \\
 \hline
 $\displaystyle X=-\frac{1}{2} g^{\mu\nu}\partial_\mu \varphi \partial_\nu \varphi$ & Kinetic term associated to the Horndeski's field.  \\
 \hline
 $\Phi$ & New measure, Eq. (\ref{DEFMED1}).  \\
 \hline
 $\psi$ &   \\
 $\chi$ & Metric perturbations, Eq. (\ref{kess5}). \\
 $\phi$ &   \\
 \hline
 $\Psi=a^3+\Phi_0$ & Auxiliary field used in subsection \ref{ejemplo}.\\
 \hline
  $\phi_N$, $\psi_N$ & Bardeen potentials.\\
 \hline
\end{tabular}
\label{tab1}
\caption{Reference to the variables used in this work.}
\end{table}

\section{The two field measure theory}\label{section:TMT}

In this Section we describe the main features of TMT, that we will use throughout this work, following the discussions presented in~\cite{Guendelman:2006wr,Cordero:2019mze}. In this theory, the action has two measures: the term with the usual metric $g_{\mu\nu}$ and an extra term with a new measure $\Phi$, in the same space-time,
\begin{equation}
\label{TMTACTION1}
S=\int \mathcal{L}_{1}\sqrt{-g}d^{4}x+\int \mathcal{L}_{2}\Phi d^{4}x.
\end{equation}
It is possible to construct the new measure using four scalar
fields, $\xi^{a}$ ($a=1,2,3,4$),
\begin{equation}
\label{DEFMED1}
\Phi=\epsilon^{\mu\nu\alpha\beta}\epsilon_{ijkl}\partial_{\mu}\xi^{i}\partial_{\nu}\xi^{j}\partial_{\alpha}\xi^{k}\partial_{\beta}\xi^{l}.
\end{equation}
In the present work we consider the Lagrangians $\mathcal{L}_{1}$, $\mathcal{L}_{2}$ to be independent from the fields $\xi^{a}$ and each of them to be independent of each other. The relations between the fields come from the equations of motion.

As shown in \cite{Guendelman:1998ms}, the variation of the action produces the equation of motion
\begin{equation}
\label{TMTEQMOV1}
A^{\mu}_{i}\partial_{\mu}\mathcal{L}_{2}=0,
\end{equation}
where
\begin{equation}
A^{\mu}_{i}=\epsilon^{\mu\nu\alpha\beta}\epsilon_{ijkl}\partial_{\nu}\xi^{j}\partial_{\alpha}\xi^{k}\partial_{\beta}\xi^{l}.
\end{equation}
Considering the relation
\begin{equation}
\label{TMTDET1}
\det\left(A^{\mu}_{i}\right)=\frac{4^{-4}}{4!}\Phi^{3}  \,,
\end{equation}
and if we assume $\Phi \neq0$, the constancy of the Lagrangian $\mathcal{L}_2$ arises.

In the following sections we will assume that the first Lagrangian contains the usual
Einstein-Hilbert action plus a matter Lagrangian which will be repeated in the second term.
In this work we will consider the matter Lagrangian to be either the
k-essence or braiding scalar field, but it can include the other matter-energy components of the Universe:
\begin{eqnarray}
    \mathcal{L}_{1} &=&
    \frac{M_p^2}{2}R + \mathcal{L}_{M}
    \\
    \mathcal{L}_{2} &=&  \mathcal{L}_{M},
\end{eqnarray}
where $\mpl = 1/\sqrt{8\pi G}$ is the reduced Planck mass ($c = \hbar = 1$).
The considerations of this section imply the constancy of $\mathcal{L}_{M}$.


\section{K-essence model perturbations}\label{section:kessence}
In this section, we will first derive the cosmological gravitational perturbations for a Universe
filled with a purely kinetic k-essence field
and calculate its sound speed in the absence
of the second measure. For this we will follow De Felice and Tsujikawa~\cite{DeFelice:2011bh}
who compute the perturbations for a Universe filled with a general Horndeski's model.
We will repeat the procedure for the model including the new measure and compare the results.

We start by considering that the action of a k-essence field $\mathcal{L}_{1}=\frac{\mpl^{2}}{2}R + K(\varphi,X)$ with $\mathcal{L}_{2}=0$:
\begin{align}
\label{kess1}
S = \int d^{4}x \: \sqrt{-g} \:\left[\frac{\mpl^{2}}{2}R \; + \; K\left(\varphi, X \right) \right].
\end{align}
The flat Friedmann equations associated with this field are given by \cite{Armendariz-Picon:1999hyi}:
\begin{eqnarray}
\label{kess2}
H^{2} &=& \frac{1}{3\mpl^{2}} \epsilon \,, \\
\label{kess3}
\frac{\ddot{a}}{a} &=& -\frac{1}{6 \mpl^{2}} \left( \epsilon + 3 P\right) \,,
\end{eqnarray}
with
\begin{eqnarray}
\label{kess4}
\epsilon_k = 2X K_{,X} - K \,, &&  P_k = K \,.
\end{eqnarray}

We will work with a flat Friedmann-Robertson-Walker spacetime plus scalar perturbations, that
has a metric of the form
\begin{equation}
\label{kess5}
ds^{2} = - \left(1 + 2 \psi \right) dt^{2} + 2  \partial_{i}\: \chi\: dt\: dx^{i} +
a^2 E_{ij}dx^idx^j+
a^2\left( 1 + 2\phi \right) d \mathbf{x}^{2} \,.
\end{equation}
The selection of a gauge for this metric usually involves choosing two of the perturbations as zero \cite{Ma:1995ey}. Here we will use the uniform field gauge where the k-essence field perturbation vanishes
($\delta\varphi = 0$) \cite{Maldacena:2002vr} and, to fix the remaining gauge freedom, also set $E_{ij}=0$.

To obtain the first order equations of motion for the perturbation fields we expand the action (\ref{kess1}) up to
second order as $S\approx S^{(0)}+ S^{(1)}+ S^{(2)}$. Using the metric (\ref{kess5})
as well as the background Friedmann equations we get the expression:
\begin{eqnarray}
\label{kess6}
S^{\:(2)} &=& \int d^{4}x a^3 \left[
\left( 2\omega_{1} \dot{\phi} -\omega_{2}\psi \right)
\frac{\partial^{2}\chi}{a^2} + \frac{\omega_{3}}{3} \psi^{2} + \frac{\omega_{4}}{a^2} \left( \partial \phi\right)^{2} - 3\omega_{1}\dot{\phi}^2
\right. \\ \nonumber && \quad \quad \quad \quad \left.
+ \left(3\omega_{2}\dot{\phi}  - 2 \omega_{1} \frac{\partial^{2} \phi}{a^2} \right) \psi \right],
\end{eqnarray}
where the $\omega_i$ coefficients are taken from Ref. \cite{DeFelice:2011bh} and we
will use them to compare with the TMT case. Here they reduce to
\begin{eqnarray}
\label{kess7}
\omega_{1}&=& \omega_{4} =  \mpl^{2},
\\
\label{kess8}
\omega_{2} &=& 2\mpl^{2} H,
\\
\label{kess9}
\omega_{3} &=& \frac{3}{2}\dot{\varphi}^{2}\left( K_{,X} + \dot{\varphi}^{2}K_{,XX}\right)- 9 \mpl^{2} H^{2} .
\end{eqnarray}

One can obtain the linear equations of motion for each scalar perturbation by varying the second order action (\ref{kess6}) with respect of each of them. In general the
expressions can be quite involved, but we can use the procedure in Ref.
\cite{DeFelice:2016ucp} to obtain the speed of the perturbations directly from the
terms in the action. In the following paragraphs we summarize the steps behind this procedure.


Considering some of the perturbation fields as Lagrange multipliers we can use their
equations of motion
and some integrations by parts, to arrive at the second order action in the general form:
\begin{align}
\label{kess10-a}
S^{(2)} = \int d^{4}x \left[ A_{ij} \dot{Q}_{i}\dot{Q}_{j} - C_{ij}\left(\partial Q_i\right)\left(\partial Q_j\right) - B_{ij}Q_{i}\dot{Q}_{j} - D_{ij}Q_{i}Q_{j}\right],
\end{align}
where $Q_i$ are the dynamical perturbation fields of the theory and we are summing over
repeated indices. In general, $A, B, C$ and $D$ will be some square matrices dependent only on
background quantities. The associated Euler-Lagrange equations become
\begin{equation}
    2A_{ij} \ddot{Q}_j +
    (2\dot A_{ij} + B_{ij} - B_{ji})\dot{Q}_j -
    2C_{ij}\partial^2Q_j +
    (2D_{ij}-\dot B_{ij}) Q_j = 0\,.
\end{equation}
In order to find the sound speed, we are interested in perturbations inside the horizon, with  $k\gg a H$. After  Fourier-transforming the fields $Q_i$ we keep only the higher terms in $\omega$ and $k$,
arriving at
\begin{equation}
    (\omega^2A_{ij}-k^2C_{ij})\widetilde Q_j(\omega,k)=0\,,
\end{equation}
that can be solved for $Q_j\neq 0$ only if the determinant of the coefficient is zero. Using
$\omega=c_sk/a$ we obtain the condition
\begin{equation}
\label{kess11}
\det\left(c_{S}^{2}A - a^{2}C \right) = 0,
\end{equation}
which can be used to obtain the speed of propagation of the perturbations in the limit
of high frequencies. If $c_s^2$ is possitive, then the perturbations propagate without growing, if it
is zero, the perturbations grow at the rate needed for the structure formation, if it is negative
the perturbations grow exponentially.

Coming back to the k-essence field, we can use the equation of motion for $\chi,\psi,\phi$,
\begin{eqnarray}
\label{kess10}
E_\chi &\to&  \partial^2(2\omega_{1} \dot{\phi} - \omega_{2} \psi) = 0\,, \\
E_{\psi} &\to& 2\frac{\omega_{3}a^{3}}{3} \psi + 3 \omega_{2}a^{3} \dot{\phi} - 2\omega_{1} a \partial ^{2} \phi - \omega_{2}a \partial^{2} \chi = 0 \,, \\
E_{\phi} &\to& -\frac{d}{dt} \left( 2\omega_{1} a\partial^{2}\chi - 6\omega_{1}a^{3}\dot{\phi}+ 3 \omega_{2}a^{3} \psi \right)   -2\omega_{4}a\partial^{2} \phi - 2 \omega_{1}a \partial^{2}\psi = 0\,,
\end{eqnarray}
to write the action in the form (\ref{kess10-a}) in terms  of  a single
perturbation field $Q_{i} =(\phi)$.
The speed of sound $c_{s}^{2}$ in this case is given by:
\begin{align}
\label{kess11-a}
c_{S}^{2} = \frac{3\left(2\omega_{1}^{2}\omega_{2}H - \omega_{2}^{2}\omega_{4} - 2\omega_{1}^{2} \dot{\omega}_{2}\right)}{\omega_{1}\left( 4\omega_{1} \omega_{3} + 9\omega_{2}^{2}\right) }.
\end{align}
Taking into account that $\partial \epsilon_k/{\partial X}= K_{,X}+2XK_{,XX}$ and substituting the values of $\omega_1, \omega_2$ and $\omega_3$ we obtained the well known result \cite{Armendariz-Picon:1999hyi}
\begin{equation}
   c_{S}^{2} = \frac{\frac{\partial P}{\partial X}}{\frac{\partial \epsilon}{\partial X}}= \frac{\partial P}{\partial \epsilon}.
\end{equation}

\subsection{k-essence and the TMT}
In the remaining of this section we repeat the previous calculation for the
k-essence field including the new measure. First we introduce the new measure field as $\Phi =
\Phi_{0}(t) + \delta \Phi(t,\mathbf{x})$, where $\Phi_{0}$ is the value of the new measure at the background level
and $\delta \Phi$ is its perturbation.
We start again with the corresponding action for the TMT case
\begin{align}
\label{kess12}
S = \int d^{4}x \sqrt{-g} \:\frac{M_{p}^{2}}{2}R  + \int d^{4}x\: \left(\sqrt{-g} + \Phi \right) \: K\left(\varphi, X \right).
\end{align}
By writing the field equations at the background level, we arrive at the energy density and pressure of the k-essence field as
\begin{eqnarray}
\label{kess13}
\epsilon &=& \dot{\varphi}^{2}K_{,X} - K + \dot{\varphi}^{2}K_{,X} \frac{\Phi_{0}}{a^{3}}, \\
\label{kess14}
P &=& K.
\end{eqnarray}
Then we expand the action (\ref{kess12}) up to second order in the perturbations fields, and use the Friedmann equations with the
new energy density and pressure to write
\begin{eqnarray}
\label{kess15}
S^{\:(2)} &=& \int d^{4}x a^3 \left[
\left( 2\omega_{1} \dot{\phi} -\omega_{2}\psi \right)
\frac{\partial^{2}\chi}{a^2} + \frac{W_{3}}{3} \psi^{2} + \frac{\omega_{4}}{a^2} \left( \partial \phi\right)^{2} - 3\omega_{1}\dot{\phi}^2
\right. \\ \nonumber && \quad \quad \quad \quad \left.
+ \left(3\omega_{2}\dot{\phi}  - 2 \omega_{1} \frac{\partial^{2} \phi}{a^2} +
\frac{W_2}{a^3}\left( 3\Phi_0 \phi -\delta \Phi \right)
\right) \psi \right],
\end{eqnarray}
 where we defined the $\omega_{i}$'s in Eqs. (\ref{kess7}),(\ref{kess8}),(\ref{kess9}) and two new coefficients appear
 \begin{eqnarray}
 \label{kess16}
 W_{2} &=& \dot{\varphi}^{2}K_{,X}, \\ \nonumber
 W_{3} &=& \omega_{3} +\frac{3}{2}\left( 3 \dot{\varphi}^{2}K_{,X}   + \dot{\varphi}^{4} K_{,XX}\right)\frac{\Phi_{0}}{a^{3}} \,.
 \end{eqnarray}

We can vary the action (\ref{kess15}) to obtain linear equations of motion for the fields, which can be written as
\begin{eqnarray}
\label{res1}
E_{\chi} &\to& \partial^2(2\omega_{1} \dot{\phi} -  \omega_{2}  \psi) = 0,
\label{echik}\\
E_{\psi} &\to& -\omega_{2} a \partial^{2} \chi + \frac{2}{3} W_{3} a^{3} \psi + 3\omega_{2}a^{3} \dot{\phi}
- 2\omega_{1} a \partial^{2} \phi + 3W_2\Phi_{0} \phi - W_2\delta\Phi = 0,\\
E_{\phi} &\to& -2 \omega_{1}\dot{a}\partial^{2}\chi - 2 \omega_{1}a \partial^{2}\dot{\chi} - 2 \omega_{4} a \partial^{2} \phi +6\omega_{1}a^{3}\ddot{\phi} + 18 \omega_{1} a^{3} H \dot{\phi} - 9 \omega_{2}a^{3} H \psi- 3\dot{\omega}_{2}a^{3}\psi - 3\omega_{2}a^{3}\dot{\psi} \label{ephik}  \\ \nonumber
&& - 2 \omega_{1} a \partial^2 \psi + 3 W_2 \Phi_{0} \psi = 0,
\\
E_{\delta \Phi} &\to& -W_2   \psi = 0\,.
\label{epsik}
\end{eqnarray}
First we note that the new measure produces a
equation of motion $E_{\delta \Phi}$ that constrains $\psi$ to be zero. In fact this relation can be obtained  from the perturbation (in the uniform gauge) of the constancy of ${\cal{L}}_2$,  since  $\delta {\cal{L}}_2 = -\frac{1}{2} K(\varphi, X)_{,X} \partial_\alpha \varphi \partial_\beta \varphi \delta g^{\alpha \beta} =-{K}_{,X} \dot\varphi ^2 \psi =0$. This relation is correctly reproduced considering the perturbation of $\Phi$. 
Instead, if we take the perturbations of the fields $\xi^a$, we can use the relation $\delta \Phi = 4 A^\mu _i \partial_\mu \delta \xi^i$, and find that the analogous  relation to Eq.(\ref{epsik}) is given by $\partial_\mu\left(K_{,X} \dot\varphi ^2 \psi \right)=0$. This last relation 
implies that $K_{,X} \dot\varphi ^2 \psi$ is a constant, and if we consider it equal to zero we recover the result of Eq.~(\ref{epsik}).

Assuming that $\phi$ can be written as
\begin{equation}
    \phi(x,t)=\int \phi(k,t)e^{ik\cdot x} d^3k, 
\end{equation}
the equation of motion for $\chi$ (\ref{echik}) implies 
\begin{equation}
    \dot \phi(k,t) = 0\,,
\end{equation}
or $\phi(k,t)=\phi(k)$. This is a known solution for the perturbation of a flat FRW Universe dominated by dust as it is the case of Cold Dark Matter (E.g. see Ref. \cite{mukhanov2005physical}). Here we obtain a similar solution but for a general k-essence model in the presence of a second measure field. It would be interesting to relate this result with the usual Cold Dark Matter approach. The solution for Cold Dark Matter, however, is obtained in a Newtonian or longitudinal gauge. In order to transform 
our results to the longitudinal gauge we use a gauge transformation relation (see Ref.~\cite{malik2009cosmological})
\begin{eqnarray}
\label{phin}
    \phi_N &=& 
    \psi + H\chi+a\frac{d}{dt}\left( 
    \frac{\chi}{a} \right) \,,
    \\\label{psin}
    \psi_N &=& - \phi - H\chi \,,
\end{eqnarray}
where the fields in the right-hand side of the previous equations correspond to the fields used in this work, with solutions $\psi=0$ and $\dot \phi=0$. From eq. (\ref{ephik})
\begin{equation}
    \frac{d}{dt}(a\chi)+a\phi=0\,.
\end{equation}
This model can reproduce the background  behaviour of a Universe filled with matter plus cosmological constant \cite{Guendelman:2012gg, Cordero:2019mze}. In the epoch dominated by matter, the scale factor $a\propto t^{2/3}$ leads to a metric perturbation $\chi=-3\phi t/5$. Substituting in equations (\ref{phin}) and (\ref{psin}) we arrive at
\begin{eqnarray}\label{unequal}
    \phi_N &=& 
    -\frac{3}{5}\phi \,,
    \\
    \psi_N &=&-\frac{3}{5}\phi \,. \nonumber
\end{eqnarray}
The Newtonian gravitational potentials (also known as Bardeen potentials) turn out to be constant and equal, as it is expected for a Universe filled with Dark Matter. In the following subsection we explore an alternative route to convince ourselves that the perturbations of this model resemble those of a matter component.

\subsection{Geodesic perturbations}
As obtained in section \ref{section:TMT} the equation of motion associated to the new measure implies the
constancy of the matter Lagrangian. In this case
\begin{equation}
    K(\varphi,X) = \rm{constant},
\end{equation}
which can be written as follows:
\begin{equation}
    \nabla_\alpha K = K_\varphi \varphi_{,\alpha} + K_{X}\nabla_\alpha X=0\,.
    \label{}
\end{equation}
We can rewrite this equation by  defining the 4-velocity of the field as
\begin{equation}
    U_\alpha \equiv \frac{\varphi_{,\alpha}}{\sqrt{2X}},
\end{equation}
then the constancy of the Lagrangian can be written as
\begin{equation}
    \left( \sqrt{2X}K_\varphi - K_X U^\beta \nabla_\beta X \right) U_\alpha
    - 2XK_X U^\beta \nabla_\beta U_\alpha
    =0\,.
    \label{}
\end{equation}
Since the two terms are linearly independent we arrive at the equations
\begin{eqnarray}
    \sqrt{2X} K_\varphi &=& K_X U^\beta \nabla_\beta X \,, \\
    U^\beta \nabla_\beta U_\alpha &=& 0\,.
    \label{}
\end{eqnarray}
The last equation corresponds to the geodesic equation that is satisfied independently
of the k-essence Lagrangian. This result was first obtained in Ref. \cite{Guendelman:2012gg} by means of a different approach
and it explains that in the procedure of the previous subsection. The fact that the field obeys the geodesic equation allows it to be a good
candidate for Dark Matter as it adequately reproduces the formation of structure in the Universe.

\section{Kinetic gravity braiding model perturbations}\label{section:braiding}

In this section, we compute the perturbation equations, the kinetic matrix, the mass eigenvalues and the sound speed for the kinetic gravity braiding models \cite{Deffayet:2010qz} plus the new measure field.
We first follow \cite{DeFelice:2011bh} calculations without the new measure in order to illustrate the procedure.

The action is given by
\begin{align}
\label{pert1}
S = \int d^{4}x \: \sqrt{-g} \:\left[\frac{R\: \mpl^{2}}{2} \; - \; G_{3}\left(X, \varphi\right) \square \varphi \right].
\end{align}
At the background level, the field equations we arrive are the
same Friedmann equations than for the k-essence field (\ref{kess2}) and (\ref{kess3}),
where now the energy density and pressure are:
\begin{eqnarray}
\label{pert4}
\epsilon_B =3\dot{\varphi}^{3}H G_{3,X} - \dot{\varphi}^{2}G_{3,\varphi} \,,
&&
P_B = -\dot{\varphi}^{2}\left( G_{3,\varphi} + \ddot{\varphi}G_{3,X}\right) \,.
\end{eqnarray}
After we expand the action up to second order in perturbations,
and use the Friedmann equations and some
integrations by parts, we obtain the analogous expression as Eq. (\ref{kess6}):
\begin{eqnarray}
\label{pert6}
S^{\:(2)} &=& \int d^{4}x a^3 \left[
\left( 2\omega_{1} \dot{\phi} -\omega_{2}\psi \right)
\frac{\partial^{2}\chi}{a^2} + \frac{\omega_{3}}{3} \psi^{2} + \frac{\omega_{4}}{a^2} \left( \partial \phi\right)^{2} - 3\omega_{1}\dot{\phi}^2
\right. \\ \nonumber && \quad \quad \quad \quad \left.
+ \left(3\omega_{2}\dot{\phi}  - 2 \omega_{1} \frac{\partial^{2} \phi}{a^2} \right) \psi \right],
\end{eqnarray}
where the $\omega_{i}$'s introduced in \cite{DeFelice:2011bh} are now:
\begin{eqnarray}
\label{pert7}
\omega_{1}&=& \omega_{4} =  \mpl^{2},
\\
\label{pert8}
\omega_{2} &=& -2 X \dot{\varphi} G_{3,X}  + 2 \mpl^{2} H,
\\
\label{pert9}
\omega_{3} &=& 18X^{2} \dot{\varphi} H G_{3,XX} + 36 X\dot{\varphi} H G_{3, X} - 6X^{2}G_{3,\varphi X} - 6XG_{3,\varphi} - 9 \mpl^{2} H^{2}.
\end{eqnarray}

The equation of motion for $\chi$ (\ref{kess10}),
will be used again to put the action in the form of (\ref{kess10-a}) and we only have one independent perturbation field $\phi$. We can compute the speed of sound using the expression (\ref{kess11}):
\begin{align}
\label{pert11}
c_{S}^{2} = 3\frac{2\omega_{1}^{2}\omega_{2}H - \omega_{2}^{2}\omega_{4}  - 2\omega_{1}^{2}\dot{\omega}_{2}}{\omega_{1}\left( 4\omega_{1} \omega_{3} + 9\omega_{2}^{2}\right) }.
\end{align}

If we now turn the attention to the case with TMT, we will arrive at a different and general result.
In this case, the action for $G_{3}(X, \varphi)\square \varphi$ with the new measure included is given by
\begin{align}
\label{pert12}
S = \int d^{4}x \:\frac{R\: \mpl^{2}}{2} \; \sqrt{-g}\: - \int d^{4}x\: \left(\sqrt{-g} + \Phi \right)  G_{3}\left(X,\varphi \right) \square \varphi.
\end{align}

The energy density and pressure for the Horndeski field $\varphi$ become:
\begin{eqnarray}
\label{pert13}
\epsilon &=&  \epsilon_{B} \left(1 + \frac{\Phi_{0}}{a^{3}}\right) + 6 G_{3} \dot{\varphi} H \frac{\Phi_{0}}{a^{3}} - \dot{\varphi}G_{3} \frac{\dot\Phi_{0}}{a^{3}} + G_{3} \ddot{\varphi} \frac{\Phi_{0}}{a^{3}}, \\
\label{pert14}
P &=&  P_{B}\left(1 + \frac{\Phi_{0}}{a^{3}}\right) - \dot{\varphi}G_{3} \frac{\dot{\Phi}_{0}}{a^{3}} - G_{3} \ddot{\varphi} \frac{\Phi_{0}}{a^{3}},
\end{eqnarray}
where $\epsilon_{B}$ and $P_{B}$ are the energy density and pressure from the kinetic gravity braiding model
(\ref{pert4}).
After we vary the action (\ref{pert12}) up to second order, and simplify it using Friedmann equations and some integration by-parts, the action can be written as:
\begin{eqnarray}
\label{pert15}
S^{(2)} &=&  \int d^{4}x a^3 \left[
\left( 2\omega_{1} \dot{\phi}   - V_{2} \psi + 3\frac{\Phi_0}{a^{3}}\dot{\varphi}G_{3}\phi
\right) \frac{\partial^{2} \chi}{a^2}
+ \frac{V_{3}}{3}\psi^{2}
+ \frac{\omega_{4}}{a^2} \left( \partial \phi \right)^{2}
- 3 \omega_{1} \dot{\phi}^{2} + F_{1} \phi^{2} \right.
\nonumber \\ && \quad \quad \quad \quad
+ \left( 3V_{2} \dot{\phi} - 2\frac{\omega_{1}}{a^2} \partial^{2} \phi
+ F_{2} \phi
  \right)\psi
\nonumber \\ && \quad \quad \quad \quad
+ \left.
\dot{\varphi}G_{3} \left(F_{3}\psi  +  3\dot{\phi}
- \dot{\psi}
- \frac{\partial^{2}\chi}{a^{2}} \right)
\frac{\delta \Phi}{a^3} \right]\,,
\end{eqnarray}
where
\begin{eqnarray}
V_{2} &=& \omega_{2} - \left(\dot{\varphi}^{3} G_{3,X}
+ \dot{\varphi} G_{3}\right) \frac{\Phi_{0}}{a^3},
\nonumber \\ \nonumber
V_{3} &=& \omega_{3}
+ \left(9H\dot{\varphi} G_{3} + 21 H \dot{\varphi}^{3}G_{3,X} + \frac{3}{2} H \dot{\varphi}^{5}G_{3,XX} + \frac{3}{2} \ddot{\varphi}G_{3} + 2 \dot{\varphi}^{2}\ddot{\varphi} G_{3,X} +  \right. \\
&& \left.  - \frac{3}{2} \dot{\varphi}^{2} G_{3, \varphi} - \frac{3\dot{\varphi}^{2}\ddot{\varphi}}{2} G_{3,X} - \frac{\dot{\varphi}^{4}}{2}G_{3,X\varphi}
\right) \frac{\Phi_{0}}{a^3}
- \frac{\dot{\varphi}^{3}}{2} G_{3,X} \frac{\dot{\Phi}_{0}}{a^3},
\nonumber \\
F_{1} &=& \frac{9}{2}\dot{\varphi}^{2} \left( G_{3,\varphi} + G_{3,X} \ddot{\varphi}  \right)\frac{\Phi_{0}}{a^3} +\frac{9}{2} \ddot{\varphi} G_{3}\frac{\Phi_{0}}{a^3} + \frac{9}{2} \dot{\varphi} G_{3} \frac{\dot{\Phi}_{0}}{a^3},
\nonumber \\
F_{2} &=& \frac{3}{a^{3}} \left( \epsilon_{B} \Phi_{0} + 6 G_{3} \dot{\varphi} H \Phi_{0} - \dot{\varphi}G_{3} \dot{\Phi}_{0} + G_{3} \ddot{\varphi} \Phi_{0} \right),
\nonumber \\
F_3 &=& - \frac{1}{\dot\varphi G_3}\left( 6H \dot{\varphi} G_{3} + 3 H \dot{\varphi}^{3} G_{3,X}  + \dot{\varphi}^{2} \ddot{\varphi} G_{3,X} + 2 \ddot{\varphi} G_{3} \right)\,,
\label{funcionesauxiliares}
\end{eqnarray}
and $\omega_{i}$ were defined by Eqs. (\ref{pert7})--(\ref{pert9}).

To obtain the speed of sound, we cancel the fields $\psi$ and $\chi$ by using their equations of motion, choosing $\phi$ and $\delta \Phi$ as the dynamical variables. The action can be written in the form (\ref{kess10-a}) with
$Q_{i} = (\phi, \delta \Phi)$ and $A$, $B$, $C$, $D$ will be $2\times2$ matrices.
To this purpose we use the equation of motion for $\chi$ that this time has contributions coming from the new measure:
\begin{align}
\label{pert16}
E_\chi \to -2\omega_{1} \dot{\phi} + V_{2}\psi -3\dot{\varphi}G_{3}\frac{\Phi_{0}}{a^3}\phi + \dot{\varphi} G_{3}\frac{\delta \Phi}{a^3} = 0.
\end{align}
The equation for $\delta \Phi$ becomes:
\begin{align}
\label{pert22}
E_{\delta \Phi} \to \dot{\varphi}G_{3}F_{3}\psi  +  3\dot{\varphi}G_{3}\dot{\phi} - \dot{\varphi}G_{3}\dot{\psi} - \dot{\varphi}G_{3}\frac{\partial^{2}\chi}{a^{2}}  = 0.
\end{align}
From the $E_{\chi}$ (\ref{pert16}), we can obtain $\psi$ as a function of $Q = (\phi, \delta\Phi)$:
\begin{equation}
\label{pert17}
\psi = P_{1}\dot{\phi} + P_{2}\phi + P_{3}\delta\Phi,
\end{equation}
where
\begin{eqnarray}
P_{1} &=& \frac{ 2\omega_{1}}{V_{2}}, \nonumber \\
P_{2} &=& \frac{3 \dot{\varphi}G_{3}}{V_{2}}\frac{\Phi_0}{a^3}, \\ \nonumber
P_{3} &=& -\frac{ \dot{\varphi}G_{3}}{a^3V_{2}}.
\end{eqnarray}

Using $E_\chi$  (\ref{pert16}) and integrating by parts
we can rewrite the action (\ref{pert15}) in terms of the perturbation fields $\phi$ and $\delta\Phi $
\begin{align}\label{brading_second}
S^{(2)} = \int d^{4}x \left( a^3\left[\frac{V_{3}}{3}P_{1}^{2}- 3 \omega_{1} + 3 V_{2} P_{1}\right] \dot{\phi}^{2} + \left[ \dot{\varphi} G_{3} P_{1}\right] \dot{\phi} \dot{\delta\Phi}  \right.
\\ \nonumber
\left. - a^3 \left[- \frac{V_{3}}{3} P_{2}^{2} - F_{1} -  F_{2}P_{2}\right]  \phi^{2} - \left[- 2\frac{a^3 V_{3}}{3}P_{2} P_{3} - a^{3} F_{2}P_{3} - \dot\varphi G_3 F_{3}P_{2} - P_{2} \frac{d}{dt} \left( \dot{\varphi}G_{3} \right)\right] \phi \delta\Phi\right.
\\ \nonumber
\left. - \left[ - \frac{a^3 V_{3}}{3} P_{3} ^{2} - \dot\varphi G_3 F_{3}P_{3} - P_{3} \frac{d}{dt}\left( \dot{\varphi} G_{3} \right) \right] \delta\Phi^{2}  - \left[- \omega_{4}a + \frac{d}{dt}\left( a \omega_{1} P_{1} \right) -2\omega_{1}a P_{2}\right] \left(\partial \phi\right)^{2} \right.
\\ \nonumber
\left. - \left[- 2 \omega_{1} a P_{3}\right] \partial \phi \partial \delta\Phi  - a^3\left[- 2\frac{V_{3}}{3}P_{1} P_{2} - 3V_{2}P_{2} - F_{2}P_{1}\right] \phi \dot{\phi} \right.
\\ \nonumber
\left.- \left[ - 2 a^3\frac{V_{3}}{3} P_{1}P_{3} - 3V_{2} P_{3}a^{3} -\dot\varphi G_3 F_{3} P_{1} - 3 \dot{\varphi}G_{3} - P_{1} \frac{d}{dt}\left( \dot{\varphi}G_{3}\right)\right] \delta\Phi \dot{\phi} \right.
\\ \nonumber
\left.
   - \left[ - \dot{\varphi} G_{3} P_{2}\right] \phi \dot{\delta\Phi} - \left[ -\dot{\varphi} G_{3} P_{3} \right] \delta\Phi \dot{\delta\Phi} \right) \,.
\end{align}
If we define $Q_i=(\phi,\delta\Phi)$ the previous action takes the form of (\ref{kess10-a}) where the matrices $A_{ij}$, $B_{ij}$,
$C_{ij}$, $D_{ij}$ can be easily read (see appendix \ref{apend}). The equation (\ref{kess11}) can be used to obtain the sound speed as:
\begin{align}
\label{pert21}
\begin{vmatrix}
c_{S}^{2}A_{11} - a^{2}C_{11} & \quad c_{S}^{2}A_{12} - a^{2}C_{12} \\
c_{S}^{2}A_{21} - a^{2}C_{21} & \quad 0
\end{vmatrix} =  0 ,
\end{align}
and implies that 
\begin{equation}
c_{S}^{2} = a^{2} \frac{C_{12}}{A_{12}} = a^{2} \frac{-\omega_{1}aP_{3}}{\dot{\varphi}G_{3}P_{1}/2} = a^{2}\frac{\omega_{1}a\frac{1}{a^{3}} \dot{\varphi} G_{3}}{\dot{\varphi}G_{3}\frac{2\omega_{1}}{2}} = 1,
\label{velocidaddelaluz}
\end{equation}
which is independent of the model \footnote{The same result is obtained if we use $Q_i=(\phi, \psi)$ instead of $Q_i=(\phi, \delta \Phi)$.}. This result arises because the new measure field imposes the constraint of a constant value for the matter Lagrangian.
This value for the sound speed makes the kinetic gravity braiding model with new measure unsuitable to represent the Dark Matter of the Universe, as the perturbations will not grow as expected from the structure formations. On the other hand, the field can still be a candidate to the Dark Energy. In order to explore further the existence of problems with the theory, in the following subsection we will investigate its possible ghost or tachyonic instabilities.

\subsection{Ghost and tachyonic instabilities}
In this section we study the ghost and tachyon instabilities of the kinetic gravity braiding model when the new measure field $\Phi$
is added. To do this we must transform the kinetic term in the perturbed action (\ref{brading_second}) to the canonical form. First we Fourier transform the action to
\begin{equation}
\label{sintransformar}
S^{(2)}= \int d^3k dt a^3 (A_{ij} \dot{Q}_i \dot{Q}_j  - B_{ij}\dot{Q}_i Q_j -\tilde{D}_{ij}Q_i Q_j) \,,
\end{equation}
were $Q=(\phi,\delta\Phi)$
and the expressions for the matrices $A$, $B$ and $D$ are written in Appendix \ref{apend}.
Comparing this action with the general form (\ref{kess10-a}) we see that the Fourier
transformation allowed us to merge the two last terms as
\begin{equation}
    C_{ij}(\partial Q_i)(\partial Q_j) + D_{ij}Q_iQ_j \to \left( -k^2C_{ij} + D_{ij} \right) Q_iQ_j \,,
\end{equation}
where we call the term in parentheses $\tilde D$.

The kinetic matrix has the structure
\begin{equation}
    A=
    \begin{pmatrix}
A_{11} & A_{12} \\
A_{12} & 0
\end{pmatrix}
\,,
\end{equation}
with eigenvalues
\begin{equation}
    \lambda_{1,2} = \frac{1}{2} (A_{11} \pm \sqrt{A_{11}^2+4A_{12}^2})\,.
\end{equation}
From this expression we see that $\lambda_1>0$ and $\lambda_2<0$ which implies that one of the resulting fields will have a negative kinetic term and therefore will be a ghost. To see this explicitly we perform the following field redefinition
\begin{eqnarray}
\phi &=& \frac{1}{\sqrt{A_{11}}}(\Psi_1 + \Psi_2) \,, \nonumber \\
\delta \Phi &=& -\frac{\sqrt{A_{11}}}{A_{12}} \Psi_2 \,,
\end{eqnarray}
which after integration by parts takes the Lagrangian to the form
\begin{equation}
\label{lag_clasic}
    \mathcal{L}=\frac{a^3}{2}
    [\dot{\Psi}_1^2 - \dot{\Psi}_2^2 + K (\dot{\Psi}_1\Psi_2 -
    \dot{\Psi}_2\Psi_1) -M_{11}\Psi_1^2 -2 M_{12}\Psi_1\Psi_2 - M_{22} \Psi_2^2] \,.
\end{equation}
where the expressions for the functions $K$ and $M_{ij}$ are written in Appendix \ref{apend}. As expected, one of the
resulting fields, in this case $\Psi_2$ has a negative kinetic term and therefore is a ghost
\cite{sbisa2014classical}, with the unwanted property that its energy is unbounded from bellow.

In order to check for tachyonic instabilities, we also need to diagonalize the mass term. To keep the kinetic part invariant
while transforming the mass term, we propose an hyperbolic rotation instead of the standard rotation \cite{DeFelice:2016ucp}
\begin{eqnarray}
\Psi_1 &=&  \cosh{\alpha} \xi_1 + \sinh{\alpha} \xi_2 \,,\nonumber \\
\Psi_2 &=& \sinh{\alpha} \xi_1 + \cosh{\alpha} \xi_2 \,,
\end{eqnarray}
with
\begin{equation}
    \tanh{2\alpha} = - \frac{2M_{12}}{M_{11}+M_{22}}\,.
\end{equation}
The Lagrangian transforms to
\begin{equation}
    \mathcal{L}=\frac{a^3}{2}
    [\dot{\xi}_1^2 - \dot{\xi}_2^2 + (K+2\dot \alpha) (\dot{\xi}_1\xi_2 -
    \dot{\xi}_2\xi_1) -\mu_1^2\xi_1^2 - \mu_2^2 \xi_2^2 ]\,,
\end{equation}
where the masses are given by
\begin{eqnarray}
\mu_1^2 &=&
\dot{\alpha}^2 + \dot{\alpha}K + \frac{1}{2} (M_{11}- M_{22})  +
\frac{1}{2}\sqrt{(M_{11} + M_{22})^2-4M_{12}^2}
\frac{|M_{11} + M_{22}|}{M_{11} + M_{22}} \,,
\nonumber
\\
\mu_2^2 &=&
- \dot{\alpha}^2 -  \dot{\alpha}K - \frac{1}{2} (M_{11}- M_{22}) +
\frac{1}{2}\sqrt{(M_{11} + M_{22})^2-4M_{12}^2}
\frac{|M_{11} + M_{22}|}{M_{11} + M_{22}} \,.
\label{masascanonicas}
\end{eqnarray}
In order to have a theory free from tachyonic instabilities both squared masses need to be positive \footnote{The former Lagrangian produces a very simple form of the Hamiltonian which will be unbounded from below if the squared masses are negatives.}. In fact, it is convenient to compare these masses with the Hubble parameter in the case when $\mu_i^2 < 0$. For $k=0$, we will have strong tachyonic instability if $-\mu_i^2 \gg H^2$ and the model is unviable. However the model is physically viable if $|\mu_i^2|\leq H^2$ and the whole stability of the system will be not destroyed by the evolution of this instability in time intervals shorter than the Hubble time \cite{DeFelice:2016ucp}. Although the expressions in terms of the original functions in the action
\eqref{pert12} are rather involved, we exemplify the application of these conditions in the following subsection in order to shed
some light on how they act in a concrete case.

\subsection{Example}\label{ejemplo}

As the expressions of the last subsection are too large to give some insight on the tachyonic stability of the
Braiding model, here we will apply the stability criteria to a particular example of cosmological interest.
We propose a Braiding field which only depends on the kinetic term  $G_3(\varphi,X)=G_3(X)$.
The Lagrangian then becomes invariant under the transformation $\varphi \to \varphi + \rm{const}$, which
leads to a conserved quantity. At the background level we can write this quantity
$I$ as:
\begin{equation}
    I=G_3(X)\left[\frac{d\Psi}{dt} - 3h\left( 2X\frac{G_{3,X}}{G_3} + 1 \right) \Psi \right]  \,,
    \label{noether}
\end{equation}
where we are using\footnote{See section V of Ref. \cite{Cordero:2019mze} for details.} $\Psi=a^3+\Phi_0$.
A second conserved quantity arises from the variation of the action \eqref{pert12} with
respect to the new measure $\Phi$ which produces a constant value for the factor $G_3(X)\square \varphi$,
at the background level this becomes
\begin{equation}
    J=G_3(X)\left( \ddot \varphi +3H \dot \varphi \right) \,.
    \label{multiplicador}
\end{equation}
The energy density and pressure from this field can be written in terms of both conserved quantities,
substituting \eqref{noether} and \eqref{multiplicador} in the expressions \eqref{pert13} and \eqref{pert14}
we obtain
\begin{eqnarray}
    \epsilon &=&  J\left( \frac{\Psi}{a^3} - 1 \right) -\frac{I\dot \varphi}{a^3}\,,\nonumber \\
    P &=& - J \frac{\Psi}{a^3} \left( \frac{2XG_{3,X}(X)}{G(X)}+1 \right) +J - \frac{I \dot \varphi}{a^3}\,.
    \label{}
\end{eqnarray}

As we are interested in a particular case, we set $I=0$ and
\begin{equation}
   G_3 = r X^\gamma
    \label{particularG} \,\, ,
\end{equation}
with $\gamma$ and $r$ constants.
Equation \eqref{noether} solves to
\begin{equation}
    \Psi = La^{3(2\gamma+1)}\,,
    \label{psisolucion}
\end{equation}
with $L$ an integration constant. This solution leads to an energy density and pressure given by
\begin{eqnarray}
    \epsilon &=&  J L a ^{6\gamma} - J\,, \nonumber \\
    P &=& - J L(2\gamma+1) a ^{6\gamma} + J \,.
    \label{densidades_totales}
\end{eqnarray}
An interesting case happens with $\gamma=-1/2$ in which the energy density looks like
a matter component plus a cosmological constant. Unfortunately this field is not suited to
behave as the dark matter at the perturbation level, as its speed of sound
\eqref{velocidaddelaluz} equals the speed of light and it is incompatible with the structure formation.

In order to study the tachyonic stability of the model we need to
compute the expressions in Eq. \eqref{masascanonicas},
which depend on the expressions in Appendix \ref{apend} and ultimately are functions of
$\Phi_0$ and $G_3(X)$. For the first one we have
from Eq. \eqref{psisolucion}
\begin{equation}
    \Phi_0 = -a^3 +  La^{3(2\gamma+1)}\,.
    \label{}
\end{equation}
For $G_3(X)$ we need to solve equation \eqref{multiplicador} with $X=1/2\dot\varphi^2$ and for
the particular Lagrangian \eqref{particularG} becomes
\begin{equation}
    J=\frac{r\dot\varphi^{2\gamma}}{2^\gamma} \left( \ddot\varphi +3H\dot\varphi \right)\,.
    \label{}
\end{equation}
This equation can be solved to
\begin{equation}
    \dot \varphi = \frac{1}{a^3} \left( \frac{2^\gamma(2\gamma+1)J}{r}\int{\frac{a^{6\gamma+2}}{H}da}
    + \tilde C \right)^{1/(2\gamma+1)}\, ,
    \label{}
\end{equation}
where $\tilde C$ is an integration constant which can be fixed by given a value to $\dot{\varphi}$ at an specific time. If we assume that the energy density from \eqref{densidades_totales} is the only component of the Universe,
the Friedmann equation becomes
\begin{equation}
    \frac{H^2}{H_0^2} = \Omega_\gamma a^{6\gamma} +1-\Omega_\gamma\,.
    \label{}
\end{equation}
The former integration constants can be written in terms of the free parameters $\Omega_\gamma$ and $\gamma$ as
\begin{eqnarray}
    J&=& 3\mpl^2H_0^2(\Omega_\gamma -1)
    \,, \nonumber\\
    L&=& \frac{\Omega_\gamma}{\Omega_\gamma -1}\,.
    \label{}
\end{eqnarray}

\begin{figure}[h]
    \centering
    \includegraphics{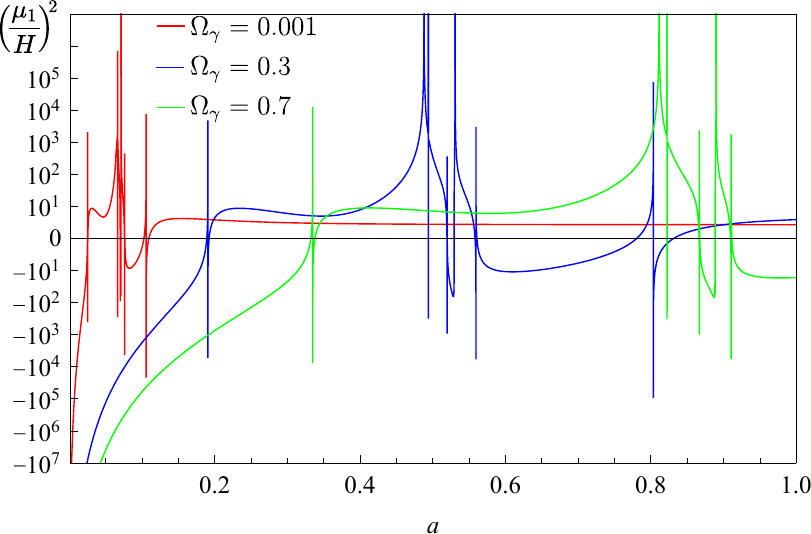}
        \includegraphics{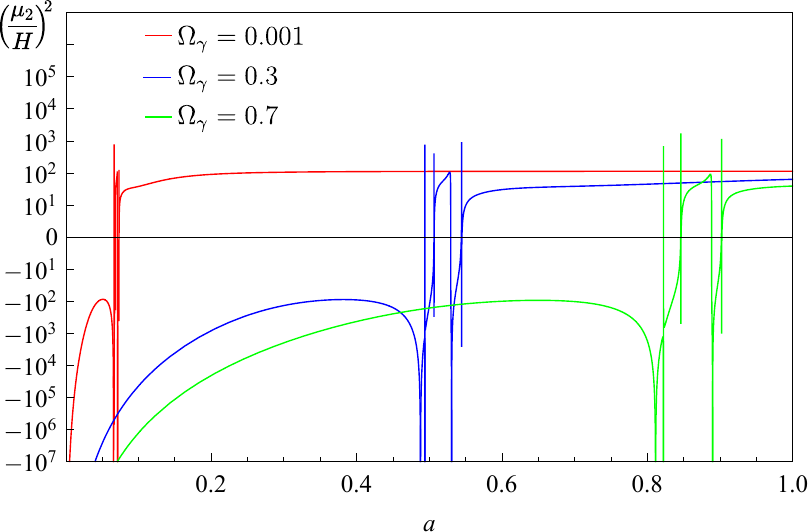}

    \caption{Evolution of the masses $\mu_1 ^2/H^2$ and $\mu_2^2/H^2$ from \eqref{masascanonicas} for the particular case studied
    in subsection \ref{ejemplo}. The first mass corresponds to the canonical fields while the second mass corresponds to the ghost field. We see that in the early Universe these masses become imaginary, corresponding to a tachyonic field. This behaviour can be modified with the inclusion of extra matter components to the Universe.}
    \label{fig:mass1}
\end{figure}

With the above equations we can study the behaviour of the masses \eqref{masascanonicas}.
Figure~\ref{fig:mass1} gives the evolution of the masses for different values of the density parameter $\Omega_\gamma$ for the case $\gamma=-1/2$. We can see changes in the sign of the squared masses, which correspond to changes in behavior from  canonical to tachyonic. For small $\Omega_\gamma$ the tachyonic behavior occurs only at the early Universe. The change from stable to tachyonic behaviour is very fast in various regions, in particular for $\Omega_\gamma =0.3$ near $a=0.52$. Furthermore, there are points where singularities are present, some of them due to a zero value for the $A_{11}$ term (see Appendix). We have such divergent behavior, for instance, for $\Omega_\gamma =0.3$ and $a\simeq0.49$ in the left panel of Figure~\ref{fig:mass1}. 

One could try a different approach by eliminating the perturbed new measure field $\delta \Phi$ of the equations of motion for the perturbations and considering instead the independent perturbations as $\phi$ and $\psi$. However, even in this case, the stability behaviour remains.  The canonical field has  a tachyonic behaviour  and the other is free from tachyonic instabilities, but it has a ghost.

If the Braiding field were intended to behave as a Dark Matter + Dark Energy, then $\Omega_\gamma \sim 0.3$, but this case present several tachyonic instabilities at different stages in the evolution of the Universe. An extension of our study will need to be performed to see if these instabilities (together with the ghost instability) can be avoided when we consider a Universe filled with baryonic matter and radiation.

\begin{table}[h]
\begin{tabular}{|l|l|l|}
\hline
\multicolumn{3}{|c|}{Velocity}
\\ \hline   & Without new measure & With new measure   \\ \hline
$K(X)$      & $c_{S}^{2} = \frac{3\left(2\omega_{1}^{2}\omega_{2}H - \omega_{2}^{2}\omega_{4}- 2 \omega_{1}^{2}\dot{\omega}_{2}\right)}{\omega_{1}\left( 4\omega_{1} \omega_{3} + 9\omega_{2}^{2}\right) }.$                    & Field follows geodesics
\\ \hline $G_{3}(X)\square \phi$ & {\begin{tabular}[c]{@{}l@{}}$c_{S}^{2} = 3\frac{2\omega_{1}^{2}\omega_{2}H - \omega_{2}^{2}\omega_{4} + 4 \omega_{2}\omega_{1}\dot{\omega}_{1} - 2\omega_{1}^{2}\dot{\omega}_{2}}{\omega_{1}\left( 4\omega_{1} \omega_{3} + 9\omega_{2}^{2}\right) }.$\\
\end{tabular}} & \multicolumn{1}{l|}{$c_{S}^{2}= 1$}

\\ \hline
\end{tabular}
\caption{The distinct results of the speed of perturbations is shown.}
\label{tab:res5}
\end{table}

\section{Conclusions}
The nature of the late time accelerating expansion of the Universe is one of the most challenging aspects of modern cosmology. One of the more appealing characteristics of Two Field Measure Theory is the possibility to have a unifying description of dark matter and dark energy.
In this paper we have studied the perturbations of k-essence and braiding fields in the context of a Two Field Measure Theory. We have found the stability characteristics and the speed of sound of the scalars fields.

In the original Horndeski theory, the speed of perturbations is a function of the field Lagrangian $K(\varphi,X)$ or $G_3(\varphi,X)$ as well as their derivatives, as it is summarized in table \ref{tab:res5}, presenting a rich variety in the speeds of sound depending on the particular Lagrangians. When the new measure $\Phi$ is added, however, both cases present a fixed behaviour. For the k-essence models we obtained that the field follows geodesics, a result already obtained in \cite{Guendelman:2012gg}. This allows the field to be a good candidate for the Dark Matter as it can grow its perturbations and form the structure of the Universe.
In contrast, kinetic gravity braiding model $G_{3}\square \phi$ has a unitary speed of perturbations. This characteristic is a consequence of the new measure field which fixed the value of the matter Lagrangian.

For the braiding models we studied further their stability conditions. We obtained that, in the presence of the new measure, there will always appear a ghost instability. The field perturbations of the braiding field plus new measure, can be mapped to a two field theory with a ghost field and a canonical field, both with complicated mass expressions written in Eq. \eqref{masascanonicas}. When the masses squared are negative we are in the presence of a tachyon instability.

To shed some light on the tachyonic stability of the Braiding model, we studied a particular example in which the field density evolves as a power law in the scale factor plus a cosmological constant. In that case, we computed the squared masses of the resulting fields, obtaining that they change their sign as the Universe evolves. When the density of the Braiding model is low, the tachyonic instability can be limited to the early Universe.

Both, ghost and tachyonic instabilities disfavour the braiding + TMT model as explanation for either dark energy or dark matter, since its energy is unbounded from below. An equally important problem for the braiding field to represent the dark matter is that its sound speed is equal to the speed of light, this prevents the perturbations to grow and form the structure of the Universe, despite its background density to evolve as dust with $1/a^3$ as can be seen in Eq. \eqref{densidades_totales} for $\gamma=-1/2$. The tachyonic and braiding instabilities might be alleviated if we had considered in our calculations a Universe with more fields besides the braiding field, however this study was outside of the scope of this work.

The dark energy nature is one of the most intriguing question in modern cosmology. The TMT theory has exciting properties. For example, any k-essence model in the context of TMT can produce a unification of dark matter and dark energy and explain the growth of structure in the Universe.

\section*{Acknowledgments}
This work was partially supported by SNI-M\'exico and CONACyT research
grant: A1-S-23238. Additionally the work of RC was partially supported
by COFAA-IPN, EDI-IPN and SIP-IPN grants: 20201841, 20210704 and 20221329.

\appendix

\section{Braided field transformations}\label{apend}
Here we write the matrix components associated to the action \eqref{sintransformar}. First for the kinetic term
$A_{ij}\dot Q_i \dot Q_j$, the matrix $A_{ij}$ has the following components
\begin{eqnarray}
    A_{11} &=&  3\mpl^2 \left( \frac{4V_3 \mpl^2}{9V_2^2} +1 \right) \,,
    \nonumber \\
    A_{12} = A_{21} &=& \frac{\dot \varphi G_3 \mpl^2}{a^3V_2} \,,
    \nonumber \\
    A_{22} &=&  0 \, ,
    \label{}
\end{eqnarray}
where the factors $V_i$ are defined in equations \eqref{funcionesauxiliares}. The second term in the action given by
$B_{ij}\dot Q_i Q_j$ has the following components for the matrix $B_{ij}$:
\begin{eqnarray}
    B_{11} &=& - G_3 \dot \varphi \frac{\Phi_0}{a^3} \left(
    \frac{4\mpl^2 V_3}{V_2^2} + 9\right) - \frac{2\mpl^2F_{2}}{V_2} \,,
    \nonumber \\
  B_{12} &=& -\frac{6X G_3^2\Phi_0}{V_2 a^6} \,,
    \nonumber \\
    B_{21} &=& \frac{2\mpl^2}{V_2 a^3} \left( \dot\varphi G_3 \left[ \frac{2V_3}{3V_2} - F_{3} \right]
               - \frac{d}{dt}[\dot\varphi G_3]\right) \,,
    \nonumber \\
    B_{22} &=& \frac{2XG_3^2}{a^6V_2} \,.
    \label{}
\end{eqnarray}
Finally the last term in the action $\tilde{D}_{ij} = -k^2C_{ij} + D_{ij}$ has the following components for the
matrix $C_{ij}$:
\begin{eqnarray}
    C_{11} &=& - \frac{\mpl^2}{a^2} + \frac{1}{a^3} \frac{d}{dt}\left( \frac{2a\mpl^4}{V_2} \right)
    - \frac{6\mpl^2\dot\varphi G_3}{V_2 a^2} \frac{\Phi_0}{a^3} \,,
    \nonumber \\
    C_{12} &=& C_{21} =  \frac{\mpl^2 \dot\varphi G_3}{a^5 V_2}\,,
    \nonumber \\
    C_{22} &=& 0 \, ,
    \label{}
\end{eqnarray}
and for the matrix $D_{ij}$:
\begin{eqnarray}
    D_{11} &=&  -\frac{3V_3 \dot\varphi^2 G_3^2}{V_2^2} \frac{\Phi_0^2}{a^6} - F_{1}
    - \frac{3F_{2} \dot \varphi G_3}{V_2} \frac{\Phi_0}{a^3}
    \nonumber \\
    D_{12} = D_{21} &=& \frac{2V_3XG_3^2\Phi_0}{V_2^2a^6} +
    \frac{F_{2}\dot\varphi G_3}{2V_2a^3} -
    \frac{3F_{3}XG_3^2\Phi_0}{V_2a^6} -
    \frac{3\dot\varphi G_3 \Phi_0}{2V_2a^6} \frac{d}{dt}\left( \dot\varphi G_3 \right) \,,
    \nonumber \\
    D_{22} &=& \frac{2XG_3^2}{V_2a^6}\left( F_{3} -\frac{2V_3}{3V_2}\right)
    +\frac{\dot\varphi G_3}{V_2a^6} \frac{d}{dt} \left( \dot\varphi G_3 \right)\,.
    \label{}
\end{eqnarray}

The function $K$ in \eqref{lag_clasic} is written in terms of the matrix components of the original action as:
\begin{equation}
    K = \frac{B_{12}-B_{21}}{2A_{12}} + \frac{\dot{A}_{11}}{A_{11}}
    - \frac{\dot{A}_{12}}{A_{12}} \,.
    \label{factork}
\end{equation}
The mass matrix which appears in Eq. \eqref{lag_clasic} is given by
\begin{eqnarray}
    M_{11} &=& \frac{1}{A_{11}} \left( \frac{1}{4} \frac{\dot A_{11}^2}{A_{11}}
    -\frac{\ddot A_{11}}{2} -\frac{\dot B_{11}}{2} +\tilde{D}_{11} -\frac{3H}{2} (\dot{A}_{11} + B_{11} )
          \right) \,,
\nonumber \\
2M_{12} &=& \left( \frac{\dot{A}_{11}}{A_{11}}
           - \frac{\dot{A}_{12}}{A_{12}} \right) \left( \frac{B_{12} - B_{21} - 2\dot{A}_{12}}{2A_{12}}
           +3H  \right)
           + \frac{\ddot{A}_{11}}{A_{11}}
           - \frac{\ddot{A}_{12}}{A_{12}}
           \nonumber \\ &&
           - \frac{\dot{B}_{11}}{A_{11}} + \frac{\dot{B}_{21} + \dot{B}_{12}}{2A_{12}}
           + \frac{2\tilde{D}_{11}}{A_{11}} - \frac{2\tilde{D}_{12}}{A_{12}}
           - \frac{3H}{2} \left(  \frac{2B_{11}}{A_{11}} - \frac{B_{21} + B_{12}}{A_{12}}\right) \,,
\nonumber \\
M_{22} &=& \left( \frac{\dot{A}_{11}}{A_{11}}
           - \frac{\dot{A}_{12}}{A_{12}} \right) \left( \frac{B_{12} - B_{21} + 2\dot{A}_{12}}{2A_{12}}
           - 3H  \right)
           - \frac{\ddot{A}_{11}}{2 A_{11}}
           - \left(\frac{\dot{A}_{11}}{2 A_{11}} \right)^2
           + \frac{\ddot{A}_{12}}{A_{12}}
           \nonumber \\ &&
           - \frac{\dot{B}_{11}}{2 A_{11}} + \frac{\dot{B}_{21} + \dot{B}_{12}}{2A_{12}}
           - \frac{\dot{B}_{22}A_{11}}{2A_{12}^2}
           + \frac{\tilde{D}_{11}}{A_{11}} - \frac{2\tilde{D}_{12}}{A_{12}}
           + \frac{\tilde{D}_{22}A_{11}}{A_{12}^2}
           \nonumber \\ &&
           - \frac{3H}{2} \left( \frac{B_{11}}{A_{11}} -\frac{B_{12}+B_{21}}{A_{12}}
           + \frac{A_{11}B_{22}}{A_{12}^2}\right) \,.
\end{eqnarray}

\bibliography{biblio}

\begin{thebibliography}{73}
\expandafter\ifx\csname natexlab\endcsname\relax\def\natexlab#1{#1}\fi
\expandafter\ifx\csname bibnamefont\endcsname\relax
  \def\bibnamefont#1{#1}\fi
\expandafter\ifx\csname bibfnamefont\endcsname\relax
  \def\bibfnamefont#1{#1}\fi
\expandafter\ifx\csname citenamefont\endcsname\relax
  \def\citenamefont#1{#1}\fi
\expandafter\ifx\csname url\endcsname\relax
  \def\url#1{\texttt{#1}}\fi
\expandafter\ifx\csname urlprefix\endcsname\relax\def\urlprefix{URL }\fi
\providecommand{\bibinfo}[2]{#2}
\providecommand{\eprint}[2][]{\url{#2}}

\bibitem[{\citenamefont{Riess et~al.}(1998)}]{Riess:1998cb}
\bibinfo{author}{\bibfnamefont{A.~G.} \bibnamefont{Riess}} \bibnamefont{et~al.}
  (\bibinfo{collaboration}{Supernova Search Team}), \bibinfo{journal}{Astron.
  J.} \textbf{\bibinfo{volume}{116}}, \bibinfo{pages}{1009}
  (\bibinfo{year}{1998}), \eprint{astro-ph/9805201}.

\bibitem[{\citenamefont{Perlmutter et~al.}(1999)}]{Perlmutter:1998np}
\bibinfo{author}{\bibfnamefont{S.}~\bibnamefont{Perlmutter}}
  \bibnamefont{et~al.} (\bibinfo{collaboration}{Supernova Cosmology Project}),
  \bibinfo{journal}{Astrophys. J.} \textbf{\bibinfo{volume}{517}},
  \bibinfo{pages}{565} (\bibinfo{year}{1999}), \eprint{astro-ph/9812133}.

\bibitem[{\citenamefont{Sahni}(2004)}]{Sahni:2004ai}
\bibinfo{author}{\bibfnamefont{V.}~\bibnamefont{Sahni}},
  \bibinfo{journal}{Lect. Notes Phys.} \textbf{\bibinfo{volume}{653}},
  \bibinfo{pages}{141} (\bibinfo{year}{2004}), \eprint{astro-ph/0403324}.

\bibitem[{\citenamefont{Alam et~al.}(2004)\citenamefont{Alam, Sahni, and
  Starobinsky}}]{Alam:2004jy}
\bibinfo{author}{\bibfnamefont{U.}~\bibnamefont{Alam}},
  \bibinfo{author}{\bibfnamefont{V.}~\bibnamefont{Sahni}}, \bibnamefont{and}
  \bibinfo{author}{\bibfnamefont{A.~A.} \bibnamefont{Starobinsky}},
  \bibinfo{journal}{JCAP} \textbf{\bibinfo{volume}{06}}, \bibinfo{pages}{008}
  (\bibinfo{year}{2004}), \eprint{astro-ph/0403687}.

\bibitem[{\citenamefont{Sahni and Starobinsky}(2006)}]{Sahni:2006pa}
\bibinfo{author}{\bibfnamefont{V.}~\bibnamefont{Sahni}} \bibnamefont{and}
  \bibinfo{author}{\bibfnamefont{A.}~\bibnamefont{Starobinsky}},
  \bibinfo{journal}{Int. J. Mod. Phys. D} \textbf{\bibinfo{volume}{15}},
  \bibinfo{pages}{2105} (\bibinfo{year}{2006}), \eprint{astro-ph/0610026}.

\bibitem[{\citenamefont{Feng et~al.}(2005)\citenamefont{Feng, Wang, and
  Zhang}}]{Feng:2004ad}
\bibinfo{author}{\bibfnamefont{B.}~\bibnamefont{Feng}},
  \bibinfo{author}{\bibfnamefont{X.-L.} \bibnamefont{Wang}}, \bibnamefont{and}
  \bibinfo{author}{\bibfnamefont{X.-M.} \bibnamefont{Zhang}},
  \bibinfo{journal}{Phys. Lett. B} \textbf{\bibinfo{volume}{607}},
  \bibinfo{pages}{35} (\bibinfo{year}{2005}), \eprint{astro-ph/0404224}.

\bibitem[{\citenamefont{Durrer and Maartens}(2008)}]{Durrer:2007re}
\bibinfo{author}{\bibfnamefont{R.}~\bibnamefont{Durrer}} \bibnamefont{and}
  \bibinfo{author}{\bibfnamefont{R.}~\bibnamefont{Maartens}},
  \bibinfo{journal}{Gen. Rel. Grav.} \textbf{\bibinfo{volume}{40}},
  \bibinfo{pages}{301} (\bibinfo{year}{2008}), \eprint{0711.0077}.

\bibitem[{\citenamefont{Bamba et~al.}(2012)\citenamefont{Bamba, Capozziello,
  Nojiri, and Odintsov}}]{Bamba:2012cp}
\bibinfo{author}{\bibfnamefont{K.}~\bibnamefont{Bamba}},
  \bibinfo{author}{\bibfnamefont{S.}~\bibnamefont{Capozziello}},
  \bibinfo{author}{\bibfnamefont{S.}~\bibnamefont{Nojiri}}, \bibnamefont{and}
  \bibinfo{author}{\bibfnamefont{S.~D.} \bibnamefont{Odintsov}},
  \bibinfo{journal}{Astrophys. Space Sci.} \textbf{\bibinfo{volume}{342}},
  \bibinfo{pages}{155} (\bibinfo{year}{2012}), \eprint{1205.3421}.

\bibitem[{\citenamefont{Bianchi and Rovelli}(2010)}]{Bianchi:2010uw}
\bibinfo{author}{\bibfnamefont{E.}~\bibnamefont{Bianchi}} \bibnamefont{and}
  \bibinfo{author}{\bibfnamefont{C.}~\bibnamefont{Rovelli}}
  (\bibinfo{year}{2010}), \eprint{1002.3966}.

\bibitem[{\citenamefont{Ratra and Peebles}(1988)}]{Ratra:1987rm}
\bibinfo{author}{\bibfnamefont{B.}~\bibnamefont{Ratra}} \bibnamefont{and}
  \bibinfo{author}{\bibfnamefont{P.~J.~E.} \bibnamefont{Peebles}},
  \bibinfo{journal}{Phys. Rev. D} \textbf{\bibinfo{volume}{37}},
  \bibinfo{pages}{3406} (\bibinfo{year}{1988}).

\bibitem[{\citenamefont{Copeland et~al.}(2006)\citenamefont{Copeland, Sami, and
  Tsujikawa}}]{Copeland:2006wr}
\bibinfo{author}{\bibfnamefont{E.~J.} \bibnamefont{Copeland}},
  \bibinfo{author}{\bibfnamefont{M.}~\bibnamefont{Sami}}, \bibnamefont{and}
  \bibinfo{author}{\bibfnamefont{S.}~\bibnamefont{Tsujikawa}},
  \bibinfo{journal}{Int. J. Mod. Phys. D} \textbf{\bibinfo{volume}{15}},
  \bibinfo{pages}{1753} (\bibinfo{year}{2006}), \eprint{hep-th/0603057}.

\bibitem[{\citenamefont{Linder}(2008)}]{Linder:2007wa}
\bibinfo{author}{\bibfnamefont{E.~V.} \bibnamefont{Linder}},
  \bibinfo{journal}{Gen. Rel. Grav.} \textbf{\bibinfo{volume}{40}},
  \bibinfo{pages}{329} (\bibinfo{year}{2008}), \eprint{0704.2064}.

\bibitem[{\citenamefont{Deffayet et~al.}(2002)\citenamefont{Deffayet, Dvali,
  and Gabadadze}}]{Deffayet:2001pu}
\bibinfo{author}{\bibfnamefont{C.}~\bibnamefont{Deffayet}},
  \bibinfo{author}{\bibfnamefont{G.~R.} \bibnamefont{Dvali}}, \bibnamefont{and}
  \bibinfo{author}{\bibfnamefont{G.}~\bibnamefont{Gabadadze}},
  \bibinfo{journal}{Phys. Rev. D} \textbf{\bibinfo{volume}{65}},
  \bibinfo{pages}{044023} (\bibinfo{year}{2002}), \eprint{astro-ph/0105068}.

\bibitem[{\citenamefont{Shtanov and Sahni}(2003)}]{Shtanov:2002mb}
\bibinfo{author}{\bibfnamefont{Y.}~\bibnamefont{Shtanov}} \bibnamefont{and}
  \bibinfo{author}{\bibfnamefont{V.}~\bibnamefont{Sahni}},
  \bibinfo{journal}{Phys. Lett. B} \textbf{\bibinfo{volume}{557}},
  \bibinfo{pages}{1} (\bibinfo{year}{2003}), \eprint{gr-qc/0208047}.

\bibitem[{\citenamefont{Amendola and Tsujikawa}(2015)}]{Amendola:2015ksp}
\bibinfo{author}{\bibfnamefont{L.}~\bibnamefont{Amendola}} \bibnamefont{and}
  \bibinfo{author}{\bibfnamefont{S.}~\bibnamefont{Tsujikawa}},
  \emph{\bibinfo{title}{{Dark Energy}: {Theory and Observations}}}
  (\bibinfo{publisher}{Cambridge University Press}, \bibinfo{year}{2015}), ISBN
  \bibinfo{isbn}{978-1-107-45398-2}.

\bibitem[{\citenamefont{Aviles et~al.}(2018)\citenamefont{Aviles,
  Rodriguez-Meza, De-Santiago, and Cervantes-Cota}}]{Aviles:2018saf}
\bibinfo{author}{\bibfnamefont{A.}~\bibnamefont{Aviles}},
  \bibinfo{author}{\bibfnamefont{M.~A.} \bibnamefont{Rodriguez-Meza}},
  \bibinfo{author}{\bibfnamefont{J.}~\bibnamefont{De-Santiago}},
  \bibnamefont{and} \bibinfo{author}{\bibfnamefont{J.~L.}
  \bibnamefont{Cervantes-Cota}}, \bibinfo{journal}{JCAP}
  \textbf{\bibinfo{volume}{11}}, \bibinfo{pages}{013} (\bibinfo{year}{2018}),
  \eprint{1809.07713}.

\bibitem[{\citenamefont{Guendelman et~al.}(2012)\citenamefont{Guendelman,
  Singleton, and Yongram}}]{Guendelman:2012gg}
\bibinfo{author}{\bibfnamefont{E.}~\bibnamefont{Guendelman}},
  \bibinfo{author}{\bibfnamefont{D.}~\bibnamefont{Singleton}},
  \bibnamefont{and} \bibinfo{author}{\bibfnamefont{N.}~\bibnamefont{Yongram}},
  \bibinfo{journal}{JCAP} \textbf{\bibinfo{volume}{11}}, \bibinfo{pages}{044}
  (\bibinfo{year}{2012}), \eprint{1205.1056}.

\bibitem[{\citenamefont{Ansoldi and Guendelman}(2013)}]{Ansoldi:2012pi}
\bibinfo{author}{\bibfnamefont{S.}~\bibnamefont{Ansoldi}} \bibnamefont{and}
  \bibinfo{author}{\bibfnamefont{E.~I.} \bibnamefont{Guendelman}},
  \bibinfo{journal}{JCAP} \textbf{\bibinfo{volume}{05}}, \bibinfo{pages}{036}
  (\bibinfo{year}{2013}), \eprint{1209.4758}.

\bibitem[{\citenamefont{Guendelman et~al.}(2013)\citenamefont{Guendelman,
  Nishino, and Rajpoot}}]{Guendelman:2013ke}
\bibinfo{author}{\bibfnamefont{E.}~\bibnamefont{Guendelman}},
  \bibinfo{author}{\bibfnamefont{H.}~\bibnamefont{Nishino}}, \bibnamefont{and}
  \bibinfo{author}{\bibfnamefont{S.}~\bibnamefont{Rajpoot}},
  \bibinfo{journal}{Phys. Rev. D} \textbf{\bibinfo{volume}{87}},
  \bibinfo{pages}{027702} (\bibinfo{year}{2013}).

\bibitem[{\citenamefont{Guendelman et~al.}(2015)\citenamefont{Guendelman,
  Nissimov, and Pacheva}}]{Guendelman:2015rea}
\bibinfo{author}{\bibfnamefont{E.}~\bibnamefont{Guendelman}},
  \bibinfo{author}{\bibfnamefont{E.}~\bibnamefont{Nissimov}}, \bibnamefont{and}
  \bibinfo{author}{\bibfnamefont{S.}~\bibnamefont{Pacheva}},
  \bibinfo{journal}{Eur. Phys. J. C} \textbf{\bibinfo{volume}{75}},
  \bibinfo{pages}{472} (\bibinfo{year}{2015}), \eprint{1508.02008}.

\bibitem[{\citenamefont{Guendelman et~al.}(2016)\citenamefont{Guendelman,
  Nissimov, and Pacheva}}]{Guendelman:2015jii}
\bibinfo{author}{\bibfnamefont{E.}~\bibnamefont{Guendelman}},
  \bibinfo{author}{\bibfnamefont{E.}~\bibnamefont{Nissimov}}, \bibnamefont{and}
  \bibinfo{author}{\bibfnamefont{S.}~\bibnamefont{Pacheva}},
  \bibinfo{journal}{Eur. Phys. J. C} \textbf{\bibinfo{volume}{76}},
  \bibinfo{pages}{90} (\bibinfo{year}{2016}), \eprint{1511.07071}.

\bibitem[{\citenamefont{Cordero et~al.}(2019)\citenamefont{Cordero, Miranda,
  and Serrano-Crivelli}}]{Cordero:2019mze}
\bibinfo{author}{\bibfnamefont{R.}~\bibnamefont{Cordero}},
  \bibinfo{author}{\bibfnamefont{O.~G.} \bibnamefont{Miranda}},
  \bibnamefont{and}
  \bibinfo{author}{\bibfnamefont{M.}~\bibnamefont{Serrano-Crivelli}},
  \bibinfo{journal}{JCAP} \textbf{\bibinfo{volume}{07}}, \bibinfo{pages}{027}
  (\bibinfo{year}{2019}), \eprint{1905.07352}.

\bibitem[{\citenamefont{Bensity et~al.}(2021)\citenamefont{Bensity, Guendelman,
  Kaganovich, Nissimov, and Pacheva}}]{Bensity:2020sfu}
\bibinfo{author}{\bibfnamefont{D.}~\bibnamefont{Bensity}},
  \bibinfo{author}{\bibfnamefont{E.~I.} \bibnamefont{Guendelman}},
  \bibinfo{author}{\bibfnamefont{A.}~\bibnamefont{Kaganovich}},
  \bibinfo{author}{\bibfnamefont{E.}~\bibnamefont{Nissimov}}, \bibnamefont{and}
  \bibinfo{author}{\bibfnamefont{S.}~\bibnamefont{Pacheva}},
  \bibinfo{journal}{Eur. Phys. J. Plus} \textbf{\bibinfo{volume}{136}},
  \bibinfo{pages}{46} (\bibinfo{year}{2021}), \eprint{2006.04063}.

\bibitem[{\citenamefont{Benisty and Guendelman}(2019)}]{Benisty:2018fja}
\bibinfo{author}{\bibfnamefont{D.}~\bibnamefont{Benisty}} \bibnamefont{and}
  \bibinfo{author}{\bibfnamefont{E.~I.} \bibnamefont{Guendelman}},
  \bibinfo{journal}{Class. Quant. Grav.} \textbf{\bibinfo{volume}{36}},
  \bibinfo{pages}{095001} (\bibinfo{year}{2019}), \eprint{1809.09866}.

\bibitem[{\citenamefont{Guendelman}(2021{\natexlab{a}})}]{Guendelman:2021nve}
\bibinfo{author}{\bibfnamefont{E.}~\bibnamefont{Guendelman}}
  (\bibinfo{year}{2021}{\natexlab{a}}), \eprint{2104.08875}.

\bibitem[{\citenamefont{Guendelman}(2021{\natexlab{b}})}]{Guendelman:2021bbr}
\bibinfo{author}{\bibfnamefont{E.~I.} \bibnamefont{Guendelman}}
  (\bibinfo{year}{2021}{\natexlab{b}}), \eprint{2105.02279}.

\bibitem[{\citenamefont{Vulfs and Guendelman}(2020)}]{Vulfs:2019xtb}
\bibinfo{author}{\bibfnamefont{T.~O.} \bibnamefont{Vulfs}} \bibnamefont{and}
  \bibinfo{author}{\bibfnamefont{E.~I.} \bibnamefont{Guendelman}},
  \bibinfo{journal}{Mod. Phys. Lett. A} \textbf{\bibinfo{volume}{35}},
  \bibinfo{pages}{2050198} (\bibinfo{year}{2020}), \eprint{1903.01792}.

\bibitem[{\citenamefont{Guendelman and
  Kaganovich}(2006{\natexlab{a}})}]{Guendelman:2006ji}
\bibinfo{author}{\bibfnamefont{E.~I.} \bibnamefont{Guendelman}}
  \bibnamefont{and} \bibinfo{author}{\bibfnamefont{A.~B.}
  \bibnamefont{Kaganovich}}, \bibinfo{journal}{Int. J. Mod. Phys. A}
  \textbf{\bibinfo{volume}{21}}, \bibinfo{pages}{4373}
  (\bibinfo{year}{2006}{\natexlab{a}}), \eprint{gr-qc/0603070}.

\bibitem[{\citenamefont{Guendelman and Kaganovich}(2013)}]{Guendelman:2012vc}
\bibinfo{author}{\bibfnamefont{E.~I.} \bibnamefont{Guendelman}}
  \bibnamefont{and} \bibinfo{author}{\bibfnamefont{A.~B.}
  \bibnamefont{Kaganovich}}, \bibinfo{journal}{Phys. Rev. D}
  \textbf{\bibinfo{volume}{87}}, \bibinfo{pages}{044021}
  (\bibinfo{year}{2013}), \eprint{1208.2132}.

\bibitem[{\citenamefont{Luty et~al.}(2003)\citenamefont{Luty, Porrati, and
  Rattazzi}}]{Luty:2003vm}
\bibinfo{author}{\bibfnamefont{M.~A.} \bibnamefont{Luty}},
  \bibinfo{author}{\bibfnamefont{M.}~\bibnamefont{Porrati}}, \bibnamefont{and}
  \bibinfo{author}{\bibfnamefont{R.}~\bibnamefont{Rattazzi}},
  \bibinfo{journal}{JHEP} \textbf{\bibinfo{volume}{09}}, \bibinfo{pages}{029}
  (\bibinfo{year}{2003}), \eprint{hep-th/0303116}.

\bibitem[{\citenamefont{Boulware and Deser}(1972)}]{Boulware:1972zf}
\bibinfo{author}{\bibfnamefont{D.~G.} \bibnamefont{Boulware}} \bibnamefont{and}
  \bibinfo{author}{\bibfnamefont{S.}~\bibnamefont{Deser}},
  \bibinfo{journal}{Phys. Lett. B} \textbf{\bibinfo{volume}{40}},
  \bibinfo{pages}{227} (\bibinfo{year}{1972}).

\bibitem[{\citenamefont{Higuchi}(1987)}]{Higuchi:1986py}
\bibinfo{author}{\bibfnamefont{A.}~\bibnamefont{Higuchi}},
  \bibinfo{journal}{Nucl. Phys. B} \textbf{\bibinfo{volume}{282}},
  \bibinfo{pages}{397} (\bibinfo{year}{1987}).

\bibitem[{\citenamefont{Guendelman and Kaganovich}(2008)}]{Guendelman:2008sv}
\bibinfo{author}{\bibfnamefont{E.~I.} \bibnamefont{Guendelman}}
  \bibnamefont{and} \bibinfo{author}{\bibfnamefont{A.~B.}
  \bibnamefont{Kaganovich}}, in \emph{\bibinfo{booktitle}{{Workshop on
  Geometry, Topology, QFT and Cosmology}}} (\bibinfo{year}{2008}),
  \eprint{0811.0793}.

\bibitem[{\citenamefont{Horndeski}(1974)}]{horndeski1974second}
\bibinfo{author}{\bibfnamefont{G.~W.} \bibnamefont{Horndeski}},
  \bibinfo{journal}{International Journal of Theoretical Physics}
  \textbf{\bibinfo{volume}{10}}, \bibinfo{pages}{363} (\bibinfo{year}{1974}).

\bibitem[{\citenamefont{Deffayet et~al.}(2011)\citenamefont{Deffayet, Gao,
  Steer, and Zahariade}}]{Deffayet:2011gz}
\bibinfo{author}{\bibfnamefont{C.}~\bibnamefont{Deffayet}},
  \bibinfo{author}{\bibfnamefont{X.}~\bibnamefont{Gao}},
  \bibinfo{author}{\bibfnamefont{D.~A.} \bibnamefont{Steer}}, \bibnamefont{and}
  \bibinfo{author}{\bibfnamefont{G.}~\bibnamefont{Zahariade}},
  \bibinfo{journal}{Phys. Rev. D} \textbf{\bibinfo{volume}{84}},
  \bibinfo{pages}{064039} (\bibinfo{year}{2011}), \eprint{1103.3260}.

\bibitem[{\citenamefont{Armendariz-Picon
  et~al.}(1999)\citenamefont{Armendariz-Picon, Damour, and
  Mukhanov}}]{Armendariz-Picon:1999hyi}
\bibinfo{author}{\bibfnamefont{C.}~\bibnamefont{Armendariz-Picon}},
  \bibinfo{author}{\bibfnamefont{T.}~\bibnamefont{Damour}}, \bibnamefont{and}
  \bibinfo{author}{\bibfnamefont{V.~F.} \bibnamefont{Mukhanov}},
  \bibinfo{journal}{Phys. Lett. B} \textbf{\bibinfo{volume}{458}},
  \bibinfo{pages}{209} (\bibinfo{year}{1999}), \eprint{hep-th/9904075}.

\bibitem[{\citenamefont{Armendariz-Picon
  et~al.}(2001)\citenamefont{Armendariz-Picon, Mukhanov, and
  Steinhardt}}]{Armendariz-Picon:2000ulo}
\bibinfo{author}{\bibfnamefont{C.}~\bibnamefont{Armendariz-Picon}},
  \bibinfo{author}{\bibfnamefont{V.~F.} \bibnamefont{Mukhanov}},
  \bibnamefont{and} \bibinfo{author}{\bibfnamefont{P.~J.}
  \bibnamefont{Steinhardt}}, \bibinfo{journal}{Phys. Rev. D}
  \textbf{\bibinfo{volume}{63}}, \bibinfo{pages}{103510}
  (\bibinfo{year}{2001}), \eprint{astro-ph/0006373}.

\bibitem[{\citenamefont{Armendariz-Picon
  et~al.}(2000)\citenamefont{Armendariz-Picon, Mukhanov, and
  Steinhardt}}]{Armendariz-Picon:2000nqq}
\bibinfo{author}{\bibfnamefont{C.}~\bibnamefont{Armendariz-Picon}},
  \bibinfo{author}{\bibfnamefont{V.~F.} \bibnamefont{Mukhanov}},
  \bibnamefont{and} \bibinfo{author}{\bibfnamefont{P.~J.}
  \bibnamefont{Steinhardt}}, \bibinfo{journal}{Phys. Rev. Lett.}
  \textbf{\bibinfo{volume}{85}}, \bibinfo{pages}{4438} (\bibinfo{year}{2000}),
  \eprint{astro-ph/0004134}.

\bibitem[{\citenamefont{Melchiorri et~al.}(2003)\citenamefont{Melchiorri,
  Mersini-Houghton, Odman, and Trodden}}]{Melchiorri:2002ux}
\bibinfo{author}{\bibfnamefont{A.}~\bibnamefont{Melchiorri}},
  \bibinfo{author}{\bibfnamefont{L.}~\bibnamefont{Mersini-Houghton}},
  \bibinfo{author}{\bibfnamefont{C.~J.} \bibnamefont{Odman}}, \bibnamefont{and}
  \bibinfo{author}{\bibfnamefont{M.}~\bibnamefont{Trodden}},
  \bibinfo{journal}{Phys. Rev. D} \textbf{\bibinfo{volume}{68}},
  \bibinfo{pages}{043509} (\bibinfo{year}{2003}), \eprint{astro-ph/0211522}.

\bibitem[{\citenamefont{Chiba et~al.}(2000)\citenamefont{Chiba, Okabe, and
  Yamaguchi}}]{Chiba:1999ka}
\bibinfo{author}{\bibfnamefont{T.}~\bibnamefont{Chiba}},
  \bibinfo{author}{\bibfnamefont{T.}~\bibnamefont{Okabe}}, \bibnamefont{and}
  \bibinfo{author}{\bibfnamefont{M.}~\bibnamefont{Yamaguchi}},
  \bibinfo{journal}{Phys. Rev. D} \textbf{\bibinfo{volume}{62}},
  \bibinfo{pages}{023511} (\bibinfo{year}{2000}), \eprint{astro-ph/9912463}.

\bibitem[{\citenamefont{Chiba}(2002)}]{Chiba:2002mw}
\bibinfo{author}{\bibfnamefont{T.}~\bibnamefont{Chiba}},
  \bibinfo{journal}{Phys. Rev. D} \textbf{\bibinfo{volume}{66}},
  \bibinfo{pages}{063514} (\bibinfo{year}{2002}), \eprint{astro-ph/0206298}.

\bibitem[{\citenamefont{Chimento and Feinstein}(2004)}]{Chimento:2003zf}
\bibinfo{author}{\bibfnamefont{L.~P.} \bibnamefont{Chimento}} \bibnamefont{and}
  \bibinfo{author}{\bibfnamefont{A.}~\bibnamefont{Feinstein}},
  \bibinfo{journal}{Mod. Phys. Lett. A} \textbf{\bibinfo{volume}{19}},
  \bibinfo{pages}{761} (\bibinfo{year}{2004}), \eprint{astro-ph/0305007}.

\bibitem[{\citenamefont{Chimento}(2004)}]{Chimento:2003ta}
\bibinfo{author}{\bibfnamefont{L.~P.} \bibnamefont{Chimento}},
  \bibinfo{journal}{Phys. Rev. D} \textbf{\bibinfo{volume}{69}},
  \bibinfo{pages}{123517} (\bibinfo{year}{2004}), \eprint{astro-ph/0311613}.

\bibitem[{\citenamefont{De-Santiago and
  Cervantes-Cota}(2011)}]{De-Santiago:2011aka}
\bibinfo{author}{\bibfnamefont{J.}~\bibnamefont{De-Santiago}} \bibnamefont{and}
  \bibinfo{author}{\bibfnamefont{J.~L.} \bibnamefont{Cervantes-Cota}},
  \bibinfo{journal}{Phys. Rev. D} \textbf{\bibinfo{volume}{83}},
  \bibinfo{pages}{063502} (\bibinfo{year}{2011}), \eprint{1102.1777}.

\bibitem[{\citenamefont{De-Santiago et~al.}(2013)\citenamefont{De-Santiago,
  Cervantes-Cota, and Wands}}]{De-Santiago:2012ibi}
\bibinfo{author}{\bibfnamefont{J.}~\bibnamefont{De-Santiago}},
  \bibinfo{author}{\bibfnamefont{J.~L.} \bibnamefont{Cervantes-Cota}},
  \bibnamefont{and} \bibinfo{author}{\bibfnamefont{D.}~\bibnamefont{Wands}},
  \bibinfo{journal}{Phys. Rev. D} \textbf{\bibinfo{volume}{87}},
  \bibinfo{pages}{023502} (\bibinfo{year}{2013}), \eprint{1204.3631}.

\bibitem[{\citenamefont{Cordero et~al.}(2017)\citenamefont{Cordero, Gonzalez,
  and Queijeiro}}]{Cordero:2016bxt}
\bibinfo{author}{\bibfnamefont{R.}~\bibnamefont{Cordero}},
  \bibinfo{author}{\bibfnamefont{E.~L.} \bibnamefont{Gonzalez}},
  \bibnamefont{and}
  \bibinfo{author}{\bibfnamefont{A.}~\bibnamefont{Queijeiro}},
  \bibinfo{journal}{Eur. Phys. J. C} \textbf{\bibinfo{volume}{77}},
  \bibinfo{pages}{413} (\bibinfo{year}{2017}), \eprint{1608.06540}.

\bibitem[{\citenamefont{Deffayet et~al.}(2010)\citenamefont{Deffayet, Pujolas,
  Sawicki, and Vikman}}]{Deffayet:2010qz}
\bibinfo{author}{\bibfnamefont{C.}~\bibnamefont{Deffayet}},
  \bibinfo{author}{\bibfnamefont{O.}~\bibnamefont{Pujolas}},
  \bibinfo{author}{\bibfnamefont{I.}~\bibnamefont{Sawicki}}, \bibnamefont{and}
  \bibinfo{author}{\bibfnamefont{A.}~\bibnamefont{Vikman}},
  \bibinfo{journal}{JCAP} \textbf{\bibinfo{volume}{10}}, \bibinfo{pages}{026}
  (\bibinfo{year}{2010}), \eprint{1008.0048}.

\bibitem[{\citenamefont{Abbott
  et~al.}(2017{\natexlab{a}})}]{LIGOScientific:2017vwq}
\bibinfo{author}{\bibfnamefont{B.~P.} \bibnamefont{Abbott}}
  \bibnamefont{et~al.} (\bibinfo{collaboration}{LIGO Scientific, Virgo}),
  \bibinfo{journal}{Phys. Rev. Lett.} \textbf{\bibinfo{volume}{119}},
  \bibinfo{pages}{161101} (\bibinfo{year}{2017}{\natexlab{a}}),
  \eprint{1710.05832}.

\bibitem[{\citenamefont{Abbott
  et~al.}(2017{\natexlab{b}})}]{LIGOScientific:2017ync}
\bibinfo{author}{\bibfnamefont{B.~P.} \bibnamefont{Abbott}}
  \bibnamefont{et~al.} (\bibinfo{collaboration}{LIGO Scientific, Virgo, Fermi
  GBM, INTEGRAL, IceCube, AstroSat Cadmium Zinc Telluride Imager Team, IPN,
  Insight-Hxmt, ANTARES, Swift, AGILE Team, 1M2H Team, Dark Energy Camera
  GW-EM, DES, DLT40, GRAWITA, Fermi-LAT, ATCA, ASKAP, Las Cumbres Observatory
  Group, OzGrav, DWF (Deeper Wider Faster Program), AST3, CAASTRO, VINROUGE,
  MASTER, J-GEM, GROWTH, JAGWAR, CaltechNRAO, TTU-NRAO, NuSTAR, Pan-STARRS,
  MAXI Team, TZAC Consortium, KU, Nordic Optical Telescope, ePESSTO, GROND,
  Texas Tech University, SALT Group, TOROS, BOOTES, MWA, CALET, IKI-GW
  Follow-up, H.E.S.S., LOFAR, LWA, HAWC, Pierre Auger, ALMA, Euro VLBI Team, Pi
  of Sky, Chandra Team at McGill University, DFN, ATLAS Telescopes, High Time
  Resolution Universe Survey, RIMAS, RATIR, SKA South Africa/MeerKAT}),
  \bibinfo{journal}{Astrophys. J. Lett.} \textbf{\bibinfo{volume}{848}},
  \bibinfo{pages}{L12} (\bibinfo{year}{2017}{\natexlab{b}}),
  \eprint{1710.05833}.

\bibitem[{\citenamefont{Abbott
  et~al.}(2017{\natexlab{c}})}]{LIGOScientific:2017zic}
\bibinfo{author}{\bibfnamefont{B.~P.} \bibnamefont{Abbott}}
  \bibnamefont{et~al.} (\bibinfo{collaboration}{LIGO Scientific, Virgo,
  Fermi-GBM, INTEGRAL}), \bibinfo{journal}{Astrophys. J. Lett.}
  \textbf{\bibinfo{volume}{848}}, \bibinfo{pages}{L13}
  (\bibinfo{year}{2017}{\natexlab{c}}), \eprint{1710.05834}.

\bibitem[{\citenamefont{Kase and Tsujikawa}(2019)}]{Kase:2018aps}
\bibinfo{author}{\bibfnamefont{R.}~\bibnamefont{Kase}} \bibnamefont{and}
  \bibinfo{author}{\bibfnamefont{S.}~\bibnamefont{Tsujikawa}},
  \bibinfo{journal}{Int. J. Mod. Phys. D} \textbf{\bibinfo{volume}{28}},
  \bibinfo{pages}{1942005} (\bibinfo{year}{2019}), \eprint{1809.08735}.

\bibitem[{\citenamefont{Kobayashi}(2019)}]{Kobayashi:2019hrl}
\bibinfo{author}{\bibfnamefont{T.}~\bibnamefont{Kobayashi}},
  \bibinfo{journal}{Rept. Prog. Phys.} \textbf{\bibinfo{volume}{82}},
  \bibinfo{pages}{086901} (\bibinfo{year}{2019}), \eprint{1901.07183}.

\bibitem[{\citenamefont{Banerjee et~al.}(2021)\citenamefont{Banerjee, Cai,
  Heisenberg, Colg\'ain, Sheikh-Jabbari, and Yang}}]{Banerjee:2020xcn}
\bibinfo{author}{\bibfnamefont{A.}~\bibnamefont{Banerjee}},
  \bibinfo{author}{\bibfnamefont{H.}~\bibnamefont{Cai}},
  \bibinfo{author}{\bibfnamefont{L.}~\bibnamefont{Heisenberg}},
  \bibinfo{author}{\bibfnamefont{E.~O.} \bibnamefont{Colg\'ain}},
  \bibinfo{author}{\bibfnamefont{M.~M.} \bibnamefont{Sheikh-Jabbari}},
  \bibnamefont{and} \bibinfo{author}{\bibfnamefont{T.}~\bibnamefont{Yang}},
  \bibinfo{journal}{Phys. Rev. D} \textbf{\bibinfo{volume}{103}},
  \bibinfo{pages}{L081305} (\bibinfo{year}{2021}), \eprint{2006.00244}.

\bibitem[{\citenamefont{Lee et~al.}(2022)\citenamefont{Lee, Lee, Colg\'ain,
  Sheikh-Jabbari, and Thakur}}]{Lee:2022cyh}
\bibinfo{author}{\bibfnamefont{B.-H.} \bibnamefont{Lee}},
  \bibinfo{author}{\bibfnamefont{W.}~\bibnamefont{Lee}},
  \bibinfo{author}{\bibfnamefont{E.~O.} \bibnamefont{Colg\'ain}},
  \bibinfo{author}{\bibfnamefont{M.~M.} \bibnamefont{Sheikh-Jabbari}},
  \bibnamefont{and} \bibinfo{author}{\bibfnamefont{S.}~\bibnamefont{Thakur}},
  \bibinfo{journal}{JCAP} \textbf{\bibinfo{volume}{04}}, \bibinfo{pages}{004}
  (\bibinfo{year}{2022}), \eprint{2202.03906}.

\bibitem[{\citenamefont{Garriga and Mukhanov}(1999)}]{Garriga:1999vw}
\bibinfo{author}{\bibfnamefont{J.}~\bibnamefont{Garriga}} \bibnamefont{and}
  \bibinfo{author}{\bibfnamefont{V.~F.} \bibnamefont{Mukhanov}},
  \bibinfo{journal}{Phys. Lett. B} \textbf{\bibinfo{volume}{458}},
  \bibinfo{pages}{219} (\bibinfo{year}{1999}), \eprint{hep-th/9904176}.

\bibitem[{\citenamefont{Nicolis et~al.}(2009)\citenamefont{Nicolis, Rattazzi,
  and Trincherini}}]{Nicolis:2008in}
\bibinfo{author}{\bibfnamefont{A.}~\bibnamefont{Nicolis}},
  \bibinfo{author}{\bibfnamefont{R.}~\bibnamefont{Rattazzi}}, \bibnamefont{and}
  \bibinfo{author}{\bibfnamefont{E.}~\bibnamefont{Trincherini}},
  \bibinfo{journal}{Phys. Rev. D} \textbf{\bibinfo{volume}{79}},
  \bibinfo{pages}{064036} (\bibinfo{year}{2009}), \eprint{0811.2197}.

\bibitem[{\citenamefont{de~Rham and Tolley}(2010)}]{deRham:2010eu}
\bibinfo{author}{\bibfnamefont{C.}~\bibnamefont{de~Rham}} \bibnamefont{and}
  \bibinfo{author}{\bibfnamefont{A.~J.} \bibnamefont{Tolley}},
  \bibinfo{journal}{JCAP} \textbf{\bibinfo{volume}{05}}, \bibinfo{pages}{015}
  (\bibinfo{year}{2010}), \eprint{1003.5917}.

\bibitem[{\citenamefont{Goon et~al.}(2011{\natexlab{a}})\citenamefont{Goon,
  Hinterbichler, and Trodden}}]{Goon:2010xh}
\bibinfo{author}{\bibfnamefont{G.~L.} \bibnamefont{Goon}},
  \bibinfo{author}{\bibfnamefont{K.}~\bibnamefont{Hinterbichler}},
  \bibnamefont{and} \bibinfo{author}{\bibfnamefont{M.}~\bibnamefont{Trodden}},
  \bibinfo{journal}{Phys. Rev. D} \textbf{\bibinfo{volume}{83}},
  \bibinfo{pages}{085015} (\bibinfo{year}{2011}{\natexlab{a}}),
  \eprint{1008.4580}.

\bibitem[{\citenamefont{Goon et~al.}(2011{\natexlab{b}})\citenamefont{Goon,
  Hinterbichler, and Trodden}}]{Goon:2011qf}
\bibinfo{author}{\bibfnamefont{G.}~\bibnamefont{Goon}},
  \bibinfo{author}{\bibfnamefont{K.}~\bibnamefont{Hinterbichler}},
  \bibnamefont{and} \bibinfo{author}{\bibfnamefont{M.}~\bibnamefont{Trodden}},
  \bibinfo{journal}{JCAP} \textbf{\bibinfo{volume}{07}}, \bibinfo{pages}{017}
  (\bibinfo{year}{2011}{\natexlab{b}}), \eprint{1103.5745}.

\bibitem[{\citenamefont{Scherrer}(2004)}]{Scherrer:2004au}
\bibinfo{author}{\bibfnamefont{R.~J.} \bibnamefont{Scherrer}},
  \bibinfo{journal}{Phys. Rev. Lett.} \textbf{\bibinfo{volume}{93}},
  \bibinfo{pages}{011301} (\bibinfo{year}{2004}), \eprint{astro-ph/0402316}.

\bibitem[{\citenamefont{Pujolas et~al.}(2011)\citenamefont{Pujolas, Sawicki,
  and Vikman}}]{Pujolas:2011he}
\bibinfo{author}{\bibfnamefont{O.}~\bibnamefont{Pujolas}},
  \bibinfo{author}{\bibfnamefont{I.}~\bibnamefont{Sawicki}}, \bibnamefont{and}
  \bibinfo{author}{\bibfnamefont{A.}~\bibnamefont{Vikman}},
  \bibinfo{journal}{JHEP} \textbf{\bibinfo{volume}{11}}, \bibinfo{pages}{156}
  (\bibinfo{year}{2011}), \eprint{1103.5360}.

\bibitem[{\citenamefont{Kimura and Yamamoto}(2011)}]{Kimura:2010di}
\bibinfo{author}{\bibfnamefont{R.}~\bibnamefont{Kimura}} \bibnamefont{and}
  \bibinfo{author}{\bibfnamefont{K.}~\bibnamefont{Yamamoto}},
  \bibinfo{journal}{JCAP} \textbf{\bibinfo{volume}{04}}, \bibinfo{pages}{025}
  (\bibinfo{year}{2011}), \eprint{1011.2006}.

\bibitem[{\citenamefont{Kimura et~al.}(2012)\citenamefont{Kimura, Kobayashi,
  and Yamamoto}}]{Kimura:2011td}
\bibinfo{author}{\bibfnamefont{R.}~\bibnamefont{Kimura}},
  \bibinfo{author}{\bibfnamefont{T.}~\bibnamefont{Kobayashi}},
  \bibnamefont{and} \bibinfo{author}{\bibfnamefont{K.}~\bibnamefont{Yamamoto}},
  \bibinfo{journal}{Phys. Rev. D} \textbf{\bibinfo{volume}{85}},
  \bibinfo{pages}{123503} (\bibinfo{year}{2012}), \eprint{1110.3598}.

\bibitem[{\citenamefont{Maity}(2013)}]{Maity:2012dx}
\bibinfo{author}{\bibfnamefont{D.}~\bibnamefont{Maity}},
  \bibinfo{journal}{Phys. Lett. B} \textbf{\bibinfo{volume}{720}},
  \bibinfo{pages}{389} (\bibinfo{year}{2013}), \eprint{1209.6554}.

\bibitem[{\citenamefont{De~Felice and Tsujikawa}(2012)}]{DeFelice:2011bh}
\bibinfo{author}{\bibfnamefont{A.}~\bibnamefont{De~Felice}} \bibnamefont{and}
  \bibinfo{author}{\bibfnamefont{S.}~\bibnamefont{Tsujikawa}},
  \bibinfo{journal}{JCAP} \textbf{\bibinfo{volume}{02}}, \bibinfo{pages}{007}
  (\bibinfo{year}{2012}), \eprint{1110.3878}.

\bibitem[{\citenamefont{Guendelman and
  Kaganovich}(2006{\natexlab{b}})}]{Guendelman:2006wr}
\bibinfo{author}{\bibfnamefont{E.~I.} \bibnamefont{Guendelman}}
  \bibnamefont{and} \bibinfo{author}{\bibfnamefont{A.~B.}
  \bibnamefont{Kaganovich}}, \bibinfo{journal}{AIP Conf. Proc.}
  \textbf{\bibinfo{volume}{861}}, \bibinfo{pages}{875}
  (\bibinfo{year}{2006}{\natexlab{b}}), \eprint{hep-th/0603229}.

\bibitem[{\citenamefont{Guendelman and Kaganovich}(1998)}]{Guendelman:1998ms}
\bibinfo{author}{\bibfnamefont{E.~I.} \bibnamefont{Guendelman}}
  \bibnamefont{and} \bibinfo{author}{\bibfnamefont{A.~B.}
  \bibnamefont{Kaganovich}}, in \emph{\bibinfo{booktitle}{{4th Alexander
  Friedmann International Seminar on Gravitation and Cosmology}}}
  (\bibinfo{year}{1998}), \eprint{gr-qc/9809052}.

\bibitem[{\citenamefont{Ma and Bertschinger}(1995)}]{Ma:1995ey}
\bibinfo{author}{\bibfnamefont{C.-P.} \bibnamefont{Ma}} \bibnamefont{and}
  \bibinfo{author}{\bibfnamefont{E.}~\bibnamefont{Bertschinger}},
  \bibinfo{journal}{Astrophys. J.} \textbf{\bibinfo{volume}{455}},
  \bibinfo{pages}{7} (\bibinfo{year}{1995}), \eprint{astro-ph/9506072}.

\bibitem[{\citenamefont{Maldacena}(2003)}]{Maldacena:2002vr}
\bibinfo{author}{\bibfnamefont{J.~M.} \bibnamefont{Maldacena}},
  \bibinfo{journal}{JHEP} \textbf{\bibinfo{volume}{05}}, \bibinfo{pages}{013}
  (\bibinfo{year}{2003}), \eprint{astro-ph/0210603}.

\bibitem[{\citenamefont{De~Felice et~al.}(2017)\citenamefont{De~Felice,
  Frusciante, and Papadomanolakis}}]{DeFelice:2016ucp}
\bibinfo{author}{\bibfnamefont{A.}~\bibnamefont{De~Felice}},
  \bibinfo{author}{\bibfnamefont{N.}~\bibnamefont{Frusciante}},
  \bibnamefont{and}
  \bibinfo{author}{\bibfnamefont{G.}~\bibnamefont{Papadomanolakis}},
  \bibinfo{journal}{JCAP} \textbf{\bibinfo{volume}{03}}, \bibinfo{pages}{027}
  (\bibinfo{year}{2017}), \eprint{1609.03599}.

\bibitem[{\citenamefont{Mukhanov}(2005)}]{mukhanov2005physical}
\bibinfo{author}{\bibfnamefont{V.}~\bibnamefont{Mukhanov}},
  \emph{\bibinfo{title}{Physical foundations of cosmology}}
  (\bibinfo{publisher}{Cambridge university press}, \bibinfo{year}{2005}).

\bibitem[{\citenamefont{Malik and Wands}(2009)}]{malik2009cosmological}
\bibinfo{author}{\bibfnamefont{K.~A.} \bibnamefont{Malik}} \bibnamefont{and}
  \bibinfo{author}{\bibfnamefont{D.}~\bibnamefont{Wands}},
  \bibinfo{journal}{Physics Reports} \textbf{\bibinfo{volume}{475}},
  \bibinfo{pages}{1} (\bibinfo{year}{2009}).

\bibitem[{\citenamefont{Sbisa}(2014)}]{sbisa2014classical}
\bibinfo{author}{\bibfnamefont{F.}~\bibnamefont{Sbisa}},
  \bibinfo{journal}{European Journal of Physics} \textbf{\bibinfo{volume}{36}},
  \bibinfo{pages}{015009} (\bibinfo{year}{2014}).

\end{thebibliography}

\end{document}